\documentclass[prl,superscriptaddress,showpacs,longbibliography,reprint]{revtex4-2}

\usepackage[colorlinks=true,linkcolor=blue,urlcolor=blue,citecolor=blue,pdfusetitle]{hyperref}
\usepackage{color}
\usepackage[utf8]{inputenc}
\usepackage[usenames,dvipsnames]{xcolor}
\usepackage{amsmath}
\usepackage{amssymb}
\usepackage{graphicx}
\graphicspath{{graphics/}}
\usepackage{epsfig}
\usepackage{dcolumn}% Align table columns on decimal point
\usepackage{bm}% bold math
\usepackage{mathrsfs}
\usepackage{multirow}
\usepackage[all]{xy}
\usepackage{pbox}
\usepackage{lipsum}
\usepackage{verbatim}
\usepackage{braket}
\usepackage{dsfont}
\usepackage{array}
\usepackage{makecell}
\usepackage{tabularx}
\usepackage{isotope}
\usepackage{xr}
\usepackage[normalem]{ulem}

\usepackage[colorlinks=true,linkcolor=blue,urlcolor=blue,citecolor=blue,pdfusetitle]{hyperref}

\begin{document}

\title{Continuous sensing and parameter estimation with the boundary time-crystal}
\author{Albert Cabot}
\affiliation{Institut f\"ur Theoretische Physik, Eberhard Karls Universit\"at T\"ubingen, Auf der
Morgenstelle 14, 72076 T\"ubingen, Germany.}
\author{Federico Carollo}
\affiliation{Institut f\"ur Theoretische Physik, Eberhard Karls Universit\"at T\"ubingen, Auf der
Morgenstelle 14, 72076 T\"ubingen, Germany.}
\author{Igor Lesanovsky}
\affiliation{Institut f\"ur Theoretische Physik, Eberhard Karls Universit\"at T\"ubingen, Auf der
Morgenstelle 14, 72076 T\"ubingen, Germany.}
\affiliation{School of Physics and Astronomy, University of Nottingham, Nottingham, NG7 2RD, UK.}
\affiliation{Centre for the Mathematics and Theoretical Physics of Quantum Non-Equilibrium Systems,
University of Nottingham, Nottingham, NG7 2RD, UK}
\begin{abstract}
A boundary time-crystal is a quantum many-body system whose dynamics is governed by the competition between coherent driving and collective dissipation. It is composed of $N$ two-level systems and features a transition between a stationary phase and an oscillatory one. The fact that the system is open allows to continuously monitor its quantum trajectories and to analyze their dependence on parameter changes. This enables the realization of a sensing device whose performance we investigate as a function of the monitoring time $T$ and of the system size $N$. We find that the best achievable sensitivity is proportional to $\sqrt{T} N$, i.e., it follows the standard quantum limit in time and Heisenberg scaling in the particle number. This theoretical scaling can be achieved in the oscillatory time-crystal phase and it is rooted in emergent quantum correlations. The main challenge is, however, to tap this capability in a measurement protocol that is experimentally feasible. We demonstrate that the standard quantum limit can be surpassed by cascading two time-crystals, where the quantum trajectories of one time-crystal are used as input for the other one. 
\end{abstract}

\maketitle

Interacting nonequilibrium quantum systems can undergo  spontaneous time-translation symmetry breaking. The phases associated with this phenomenon are called time-crystals \cite{Sacha2018,Else2020,Zaletel2023} and can be observed both in driven Hamiltonian systems  and in dissipative scenarios. In spite of a variety of possible mechanisms underlying their emergence, a common feature is the presence of asymptotic oscillations in some  observable of the system \cite{Sacha2018,Else2020,Zaletel2023}. Such oscillations either break a discrete time-symmetry, by displaying a period which is a multiple of the driving period \cite{Else2016,Khemani2016,Yao2017,Gong2018,Wang2018,Gambetta2019,Lazarides2020,Riera2020,Zhu2019,Chinzei2020,Kessler2021,Tuquero2022,Sarkar2022,Cabot2022b}, or a continuous time-symmetry \cite{Iemini2018,Tucker2018,Buonaiuto2021,Lledo2020,Carollo2022,Seibold2020,Buca2019,Buca2019b,Booker2020,Prazeres2021,Piccitto2021,Louren2022,Hajdusek2022,Krishna2023,Mattes2023}, by approaching a limit cycle under time-independent driving \cite{Strogatz2018}. The boundary time-crystal (BTC) is a paradigmatic example of (continuous) dissipative time-crystal \cite{Iemini2018}. It manifests in collective spin systems, in which the interplay between  driving and dissipation leads to an oscillatory phase \cite{Iemini2018,Carollo2022}. Various time-crystal phases have been reported in open quantum systems \cite{Gong2018,Wang2018,Gambetta2019,Lazarides2020,Riera2020,Zhu2019,Chinzei2020,Kessler2021,Tuquero2022,Sarkar2022,Tucker2018,Lledo2020,Seibold2020,Buca2019,Buca2019b,Booker2020,Prazeres2021,Piccitto2021,Hajdusek2022,Cabot2022b,Krishna2023,Mattes2023}, including recent experimental observations in atom-cavity setups \cite{Kessler2021,Kongkhambut2022}. \\
\begin{figure}[t!]
 \centering
 \includegraphics[width=\columnwidth]{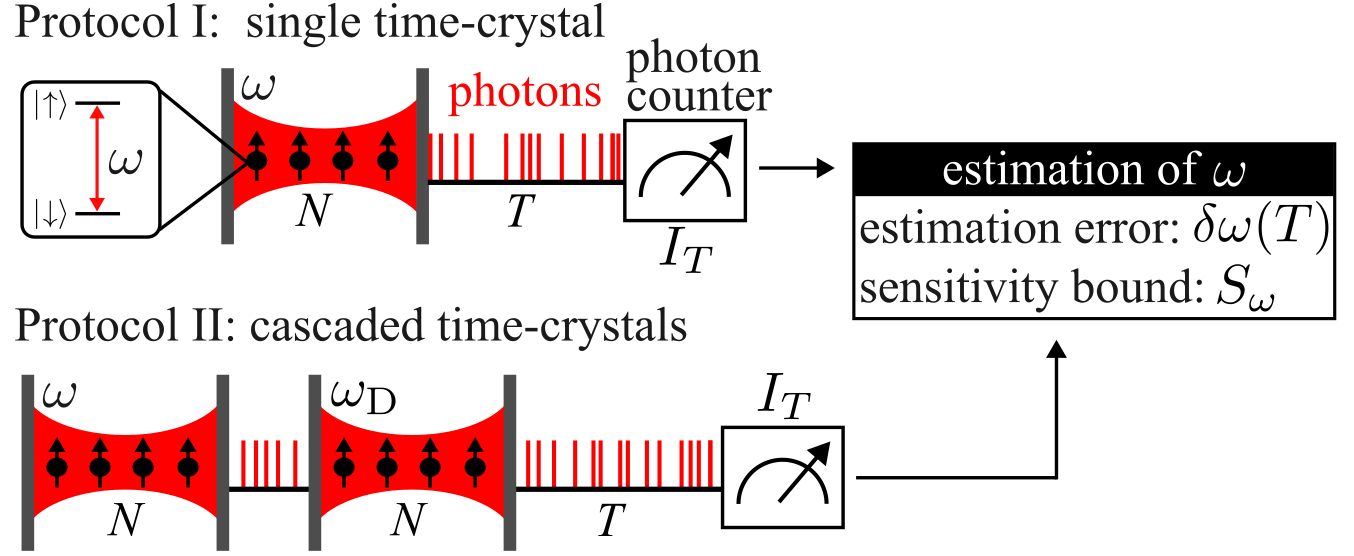}
 \caption{{\bf Continuous time-crystal sensor.} We consider $N$ two-level systems subject to collective dissipation and driven with a Rabi frequency $\omega$.
This can be realized both with cavity \cite{Kessler2021,Kongkhambut2022} or free space implementations \cite{Ferioli2022}. We consider two sensing protocols, a single time-crystal (protocol I) and a two cascaded time-crystals (protocol II), where one system is driven with Rabi frequency $\omega$ and the other with $\omega_\mathrm{D}$. In both cases the parameter $\omega$ is estimated by measuring the photon output intensity $I_T$ over a time interval of length $T$ or in the stationary limit ($T\rightarrow\infty$). The estimation error $\delta \omega(T)$ --- and whether it saturates the theoretical sensitivity bound $S_\omega$ --- depends on the observation time and the protocol.}
 \label{fig_cartoon}
\end{figure}
Dissipative time-crystals provide an example of collective phenomena in which dissipation plays a constructive role. This contrasts its more archetypal role in which dissipation simply  destroys quantum correlations, hence being usually detrimental for, e.g., quantum metrology protocols \cite{Degen2017,Braun2018}. Alternative quantum metrology approaches try to harness nonequilibrium phenomena in order to enhance the sensitivity of parameter estimation \cite{Ilias2022}. For instance, this is the case of protocols exploiting dissipative phase transitions \cite{Macieszczak2016,Fernandez2017,Heugel2019,Garbe2020,Ilias2022,Gandia2023,Pavlov2023,Montenegro2023,Ilias2023}.  Another key idea is to exploit the information contained in the emissions of open quantum systems via continuous monitoring protocols \cite{Ilias2022,Gambetta2001,Kiilerich2014,Albarelli2017,Albarelli2018,Shankar2019,Albarelli2020,Rossi2020,Nurdin2022}. Bounds to the fundamental sensitivity achievable by these {\it continuous sensors} have been derived \cite{Gammelmark2014,Catana2015}. In general, they are difficult to saturate with bare photocounting or homodyne detection protocols [cf.~Protocol I in Fig.~\ref{fig_cartoon}]. However, strategies have been developed in order to improve the sensitivity of continuous sensors \cite{Gammelmark2013,Kiilerich2016,Fallani2022},  as the use of auxiliary systems \cite{Yang2022,Godley2023}. Connecting to the latter idea, Ref. \cite{Yang2022} provides a general protocol based on cascading the output \cite{Gardiner1993,Carmichael1993,Stannigel2012} of the sensor to a replica system that is continuously monitored [cf.~Protocol II in Fig.~\ref{fig_cartoon}], which may enable the saturation of the fundamental bound. 

In this work we show how dissipative time-crystals can be exploited for sensing applications, reaching a sensitivity which can surpass the standard quantum limit. Recent works have studied BTCs from the perspective of critically enhanced sensing \cite{Pavlov2023,Montenegro2023}, focusing on properties of the system alone, while Ref.~\cite{Iemini2023} considered a discrete time-crystal for sensing time-dependent fields. Here, we instead assess the performance of BTCs as continuous sensors. The rationale is that time-crystal oscillations clearly manifest in the photocounting and homodyne detection signals even for finite sizes \cite{Cabot2022} and that this output is readily accessible in experiments. Firstly, we analyze the fundamental (theoretical) bound on the achievable sensitivity with these devices. Subsequently, we consider two different sensing protocols, based on photocounting experiments, see Fig.~\ref{fig_cartoon}. Protocol I entails the direct monitoring of the  signal while Protocol II relies on indirect monitoring of the photocurrent through a cascaded replica of the BTC, acting as a decoder, in the spirit of Ref.~\cite{Yang2022}. The time-crystal phase offers an enhanced sensitivity bound which scales linearly with the particle number $N$. This theoretical sensitivity cannot be achieved by Protocol I. Protocol II, on the other hand, allows to achieve the scaling $\sim N^{0.80}$, thus surpassing  the standard quantum limit \cite{Ilias2022}.

{\it The model. --} The BTC is composed of $N$ spin-$1/2$ particles and described by the master equation ($\hbar=1$)
\begin{equation}\label{ME_model1}
\begin{split}
\partial_t \hat{\rho}=-i\omega[ \hat{S}_\mathrm{x},\hat{\rho}]+\kappa\mathcal{D}[\hat{S}_-]\hat{\rho}\, , 
\end{split}
\end{equation}
with $\mathcal{D}[\hat{{O}}]\hat{\rho}=\hat{O}\hat{\rho}\hat{O}^\dagger-\{\hat{O}^\dagger\hat{O},\hat{\rho}\}/2$ and $\hat{\rho}$ being the state of the system. We further defined $\hat{S}_\mathrm{\alpha}=\frac{1}{2}\sum_{j=1}^N \hat{\sigma}_\mathrm{\alpha}^{(j)}$ ($\alpha=\mathrm{x,y,z}$), with $\hat{\sigma}_\mathrm{\alpha}^{(j)}$ being the Pauli matrices and $\hat{S}_\pm=\hat{S}_\mathrm{x}\pm i\hat{S}_\mathrm{y}$. Eq.~\eqref{ME_model1} thus encodes collective spin decay with rate $\kappa$ and a (resonant) driving with Rabi frequency $\omega$. It preserves the total angular momentum, and we focus throughout on the fully symmetric sector. The above model was introduced in the context of cooperative resonance fluorescence, see, e.g., Refs.~\cite{Agarwal1977,Narducci1978,Drummond1978,Carmichael1980}, and was recently recognized as a simple dissipative model displaying a time-crystal phase \cite{Iemini2018}.

For collective systems such as that of Eq.~\eqref{ME_model1}, it is customary to rescale the collective decay rate $\kappa$ by the system size in order to enforce a well-defined thermodynamic limit \cite{benatti2018,Carollo2022}. However, we focus on  finite-size systems and thus  consider a $N$-independent  rate, which further allows for a closer connection with experiments \cite{Ferioli2022}.  The system displays a stationary regime, characterized by fast relaxation to the stationary state, and an oscillatory regime, featuring long-lived oscillations \cite{Carmichael1980}. The {\it quality factor} of the oscillations increases with system size \cite{Carmichael1980} and diverges in the thermodynamic limit, where the system approaches the time-crystal phase \cite{Iemini2018,Carollo2022}. The two regimes are sharply separated by a critical Rabi frequency, $\omega_c=\kappa N/2$, as system size increases \cite{Carmichael1980}, the long-lived oscillations emerging for $\omega>\omega_c$.

{\it Continuous sensing. --} Our goal is to exploit the above system as a continuous sensor for estimating the Rabi frequency $\omega$. To this end, we consider sensing protocols based on photocounting, so that the quantity that is measured is the time-integrated photon count or output intensity $I_T$
up to a measurement time $T$. The latter is defined as  $I_T=\frac{1}{T}\int_0^T dN(t)$,
where $dN(t)$ is a random variable, taking the value $1$ when a photon is detected at time $t$ and $0$ otherwise \cite{Wiseman2009}, with average value given by $\mathbb{E}[dN(t)]=\kappa dt{\rm Tr}[\hat{S}_+\hat{S}_-\hat{\rho}]$. Here, $\mathbb{E}[\cdot]$ represents the average over all possible realizations of the photocounting process \cite{Wiseman2009}. A relevant figure of merit for the sensitivity of a protocol is the {\it estimation error}, or error propagation formula, which, for the above-introduced output intensity, reads 
\begin{equation}\label{eq_error}
\delta \omega(T)=\sqrt{\mathbb{E}[I_T^2]-\mathbb{E}[I_T]^2}\,{\left|\frac{\partial \mathbb{E}[I_T]}{\partial \omega}\right|}^{-1}\, .    
\end{equation}
The first term on the right-hand side of the above equation is the standard deviation of the measurement, while the second one represents the susceptibility of the intensity on $\omega$. In practice, the estimation error is the inverse of the signal to noise ratio. 
The {\it quantum Fisher information of the system and emission field} (QFI), $F_\mathrm{E}$, provides a lower bound for continuous sensing protocols \cite{Macieszczak2016},
\begin{equation}\label{bound}
\delta \omega(T) \geq [\sqrt{F_\mathrm{E}(\omega,T)}]^{-1}\, ,
\end{equation}
which applies to {\it any} protocol exploiting the information provided by the joint system and output state \cite{Gammelmark2014,Macieszczak2016,Yang2022}. 

Here, we focus on a long measurement time limit, in which the system is described by its stationary state $\hat{\rho}_\mathrm{ss}$. 
Since $I_T$ obeys a large deviation principle \cite{Garrahan2010,Carollo2018}, the time-integrated intensity tends to its mean \cite{Wiseman2009},  $\lim_{T\to\infty} I_T= \kappa\text{Tr}[ \hat{S}_+\hat{S}_- \hat{\rho}_\mathrm{ss}]$,
and the standard deviation scales, away from phase transitions, as
$\sqrt{(\mathbb{E}[I^2_T]-\mathbb{E}[I_T]^2)}\sim \overline{\sigma_{I_T}}/\sqrt{T}$. 
As a consequence, the estimation error asymptotically behaves as
\begin{equation}\label{error_rate}
\delta \omega \sim \frac{\overline{\delta\omega}}{\sqrt{T}}\, . 
\end{equation}
Both the prefactor for the standard deviation $\overline{\sigma_{I_T}}$ and the one for the estimation error $\overline{\delta\omega}$ are time-independent quantities (see also Supplemental Material \cite{SM}). The QFI scales linearly with time \cite{Gammelmark2014,SM}, and thus 
\begin{equation}\label{sensitivity_bound}
\sqrt{F_\mathrm{E}(\omega,T)}\sim \mathcal{S}_\omega \sqrt{T}\, .
\end{equation}
The quantity $\mathcal{S}_\omega$ in the above equation is thus the (theoretical) {\it sensitivity bound} for the estimation, i.e.,
$ (\overline{\delta \omega})^{-1}\leq \mathcal{S}_\omega$.

\begin{figure}[t!]
 \centering
 \includegraphics[width=1\columnwidth]{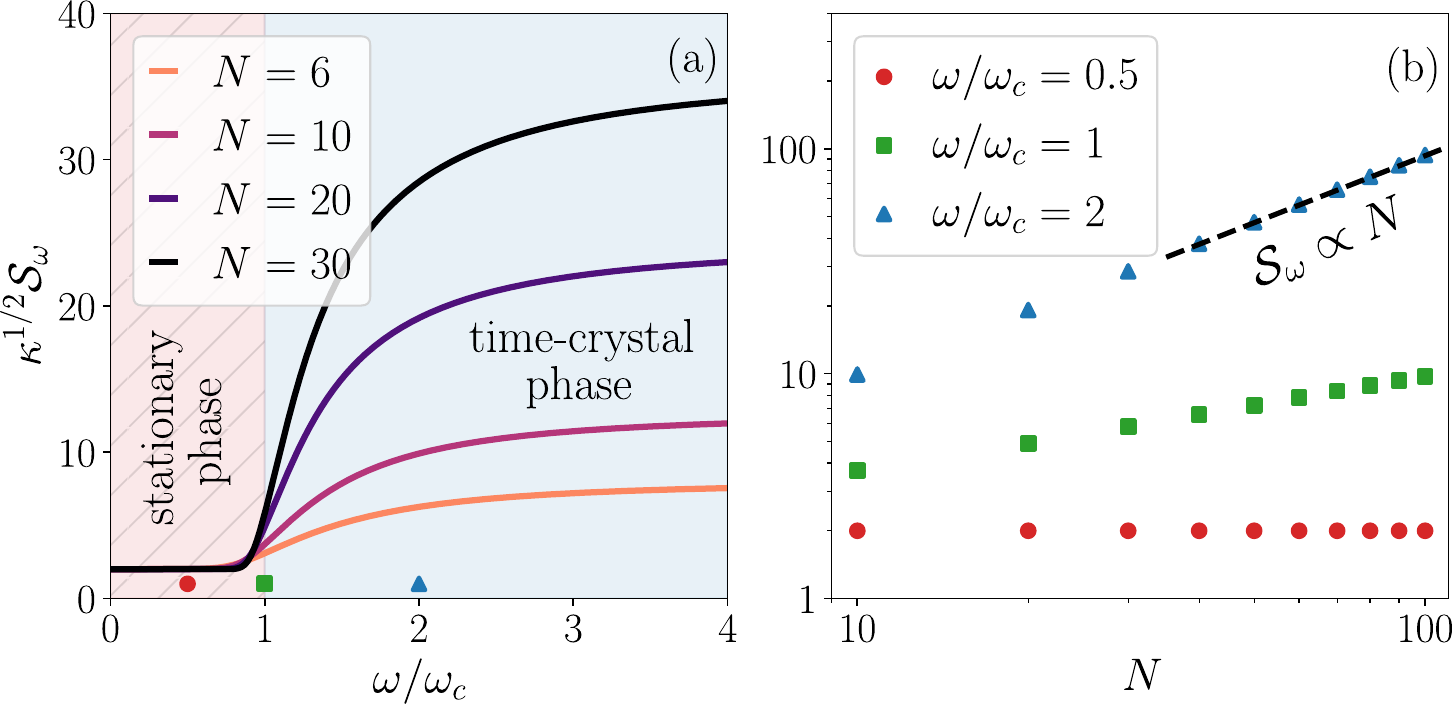}
 \caption{{\bf Sensitivity bound.} (a) Sensitivity bound $\mathcal{S}_\omega$ [cf.~Eq.~\eqref{sensitivity_bound}], as a function of the Rabi frequency $\omega$. (b) Sensitivity bound $\mathcal{S}_\omega$ as a function of the system size $N$ for different values of the Rabi frequency. The black dashed line is a fit of the six points with largest size, yielding $\mathcal{S}_\omega\propto N^{1.0}$.}
 \label{fig_QFI}
\end{figure}

{\it Fundamental bound on sensitivity. --} In Fig.~\ref{fig_QFI} we analyze the theoretically achievable bound on the sensitivity, $S_\omega$. Deep inside the stationary phase this bound is constant, $\mathcal{S}_\omega\approx 2/\sqrt{\kappa}$, i.e., it does not depend on $\omega$. This result is obtained via  a Holstein-Primakoff approach (HP) (see \cite{SM}), which shows that  spins organize in a large displaced state. Its fluctuations are annihilated by the dissipator and hence they do not contribute to the properties of the emitted light at leading order. The QFI is in this case dominated by the amplitude of the coherent displacement.

In the vicinity of $\omega_c$ we observe a sharp crossover with the sensitivity bound becoming larger in the oscillatory phase ($\omega/\omega_c>1$). This feature becomes more and more pronounced the larger the  system size. The sensitivity in the time-crystal phase indeed grows linearly, $\mathcal{S}_\omega\propto N$ [see Fig.~\ref{fig_QFI}(b)], well into the time-crystal regime (while at criticality the scaling is sublinear). From Fig.~\ref{fig_QFI}(a) it is also clear that for fixed system size $N$ the sensitivity bound saturates upon increasing $\omega/\omega_c$. This shows that increasing the system size and increasing the Rabi frequency \emph{do not} have the same effect on the sensitivity. Therefore, the $N$-dependency of the bound in the time-crystal phase is a genuine many-body effect. Notice that  $\mathcal{S}_\omega\propto N$ corresponds to a $N^2$-scaling for the QFI, meaning that the time-crystal phase theoretically offers a Heisenberg limited sensitivity in the number of particles \cite{Ilias2022}.

The system size enhancement of the sensitivity can be understood from the properties of the emergent many-body oscillations. As shown in Ref.~\cite{Cabot2022}, these oscillations translate directly into the photocounting signal, manifesting as an oscillatory detection signal. Increasing system size with fixed $\omega/\omega_c$  has a twofold effect. First, the quality factor of the oscillations increases linearly with $N$ \cite{Carmichael1980,Buonaiuto2021,Carollo2022}, and thus the oscillations in the emitted field have an increasingly better defined frequency. Second, the amplitude of the oscillations also increases with system size (as more atoms emit synchronously), and thus the signal stands out more clearly from background noise \cite{Cabot2022}. The combination of these effects creates stronger correlations between the system and the emitted field, which are a known source for enhancement of the QFI  \cite{Gammelmark2014,Macieszczak2016,Yang2022}. An indirect signature of such strong correlations is the fact that the reduced state of the system in the time-crystal phase is close to the maximally mixed state \cite{Hannukainen2018,Cabot2022}, which witnesses that the total state of system and emission field is highly correlated. Interestingly, the monitored state of the system in the time-crystal phase also displays multipartite entanglement \cite{Passarelli2023}. In the following, we explore whether sensitivities close to the bound can be achieved using sensing protocols based on photocounting.

\begin{figure}[t!]
 \centering
 \includegraphics[width=1\columnwidth]{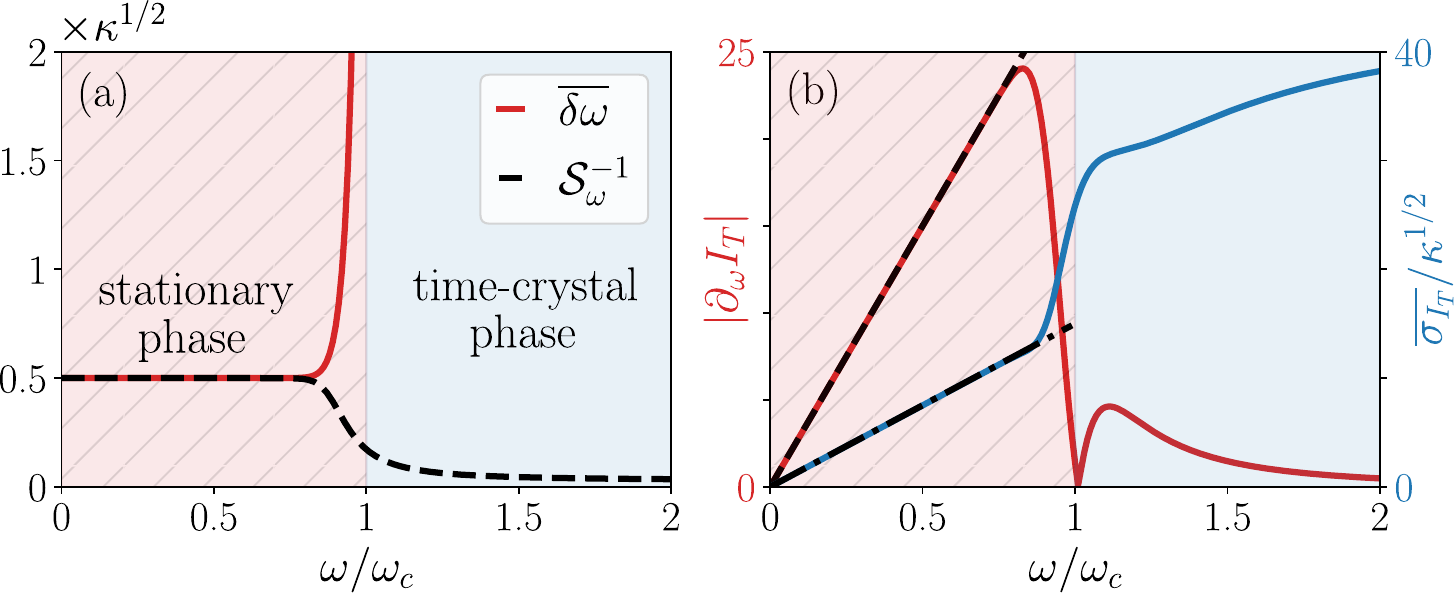}
 \caption{{\bf Protocol I.} (a) Red solid line: estimation error  $\overline{\delta \omega}$ [cf.~Eq.~\eqref{error_rate}], as a function of the Rabi frequency $\omega$ for $N=30$. Black-dashed line: inverse sensitivity bound $\mathcal{S}_\omega^{-1}$ [Eq. (\ref{sensitivity_bound})]. These quantities are plotted in units of $\sqrt{\kappa}$. (b) Left axis (red): absolute value of the derivative of the stationary intensity, as a function of the Rabi frequency. Right axis (blue): prefactor of the scaling of the standard deviation $\overline{\sigma_{I_T}}$, as a function of the Rabi frequency. Black broken lines correspond to analytical results for the stationary phase \cite{SM}.}
 \label{fig_protocol_1}
\end{figure}

{\it Protocol I. --} The first protocol we consider is based on counting the number of emitted photons in a time window $T$. For long measurement times, a single (ideal) measurement run yields the stationary state intensity, while its standard deviation can be systematically studied using large deviations \cite{SM}. In Fig.~\ref{fig_protocol_1}(a) we present the estimation error for this protocol. The smallest estimation errors are attained in the stationary phase. Indeed at the transition the estimation error  increases steeply and in the time-crystal phase it assumes values which are much larger than the bound $S_\omega$. The ``bad" performance under this measurement protocol can be understood from the data shown in Fig.~\ref{fig_protocol_1}(b). There we observe that in the time-crystal phase the derivative of the intensity  with $\omega$ diminishes significantly while the standard deviation displays the opposite trend.

\begin{figure*}[t!]
 \centering
 \includegraphics[width=2.1\columnwidth]{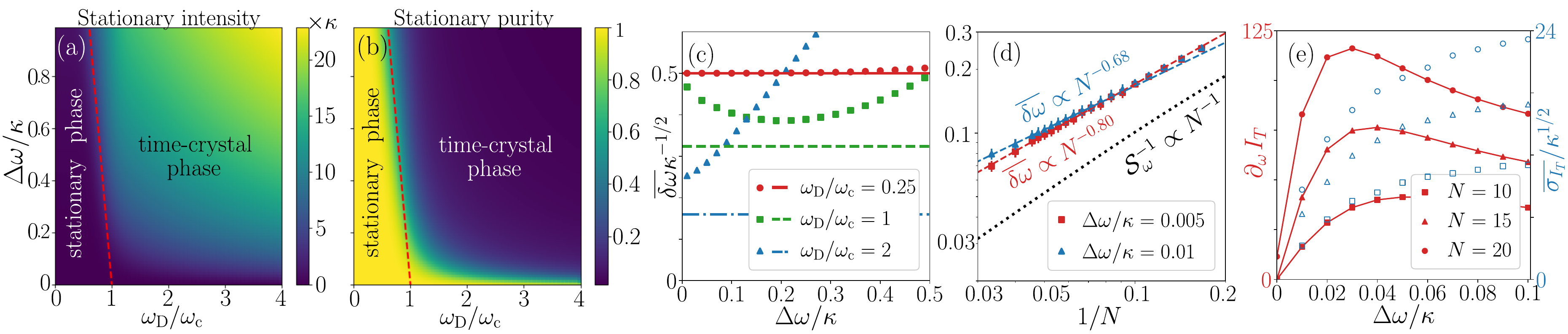}
 \caption{{\bf Protocol II.} (a) Stationary intensity of the cascaded system  varying $\Delta \omega=\omega-\omega_\mathrm{D}$ and $\omega_\mathrm{D}$. At the line $\Delta\omega=0$ the emitted intensity is zero. This quantity is plotted in units of $\kappa$. (b) Purity of the stationary state of the cascaded system. In panels (a), (b) the red dashed line corresponds to the mean-field transition line. In both cases  $N=10$. (c) Estimation error $\overline{\delta\omega}$ [cf.~Eq.~\eqref{error_rate}] varying the Rabi frequency difference between the spin systems $\Delta\omega$, and for three different points in the phase diagram. Color lines correspond to the inverse of the sensitivity bound $\mathcal{S}_\omega^{-1}$ [cf.~Eq.~\eqref{sensitivity_bound}]  displayed in Fig.~\ref{fig_QFI}(a). In this panel we have considered $N=6$. (d) Color points: estimation error $\overline{\delta\omega}$ varying system size  $N\in[6,30]$ for   $\omega_\mathrm{D}/\omega_c=2$. The black dotted line corresponds to the sensitivity bound $\mathcal{S}_\omega^{-1}$ displayed in Fig.~\ref{fig_QFI}(b). The color dashed lines correspond to fits for the eight data points with largest size, yielding the exponential laws depicted in the figure (see \cite{SM} for more details).  (e) Left (red) axis: derivative of the (stationary) emitted intensity  with respect to $\omega$  as a function of $\Delta \omega$. Right (blue) axis: prefactor for the standard deviation of the output intensity, as a function of $\Delta \omega$. In this panel $\omega_\mathrm{D}/\omega_\mathrm{c}=2$.}
 \label{fig_protocol_2}
\end{figure*}

Interestingly, simple photocounting saturates the sensitivity bound in the stationary phase. This can be understood by applying the HP approximation to methods from large deviation theory, which allows us to obtain an analytical approximation for the error \cite{SM}: $\overline{\delta \omega}\approx0.5\sqrt{\kappa}$. This value agrees with the numerical results of Fig.~\ref{fig_protocol_1}(b) (see black-broken lines). The physical interpretation is the same as for the QFI, i.e.,  to leading order the emitted light displays the same properties as a state with large coherent amplitude.

{\it Protocol II. --} In the second protocol, inspired by Ref.~\cite{Yang2022}, the emission of the system of interest, the {\it sensor}, is cascaded \cite{Gardiner1993,Carmichael1993} into an auxiliary system with same degrees of freedom, referred to as {\it decoder} (see Fig.~\ref{fig_cartoon}). This protocol builds on the emergence of dark states in the cascaded system, a situation explored in detail in Ref.~\cite{Stannigel2012}. Optimal sensing is then achieved close to the parameters in which the dark state is present \cite{Yang2022}. We consider two BTCs in cascaded configuration: all output light of the sensor (system 1) is fed into the decoder (system 2), while no light of decoder returns to the sensor \cite{Gardiner1993,Carmichael1993,Stannigel2012}. Protocol II is described by the master equation:
\begin{equation}\label{ME_model2}
\partial_t\hat{\rho}=-i[\omega \hat{S}_\mathrm{x}^{(1)}+\omega_\mathrm{D} \hat{S}_\mathrm{x}^{(2)}+\hat{H}_\mathrm{c},\hat{\rho}]
+\kappa\mathcal{D}[\hat{J}_-]\hat{\rho}.
\end{equation}
where $\hat{H}_\mathrm{c}=-i\kappa(\hat{S}_+^{(2)}\hat{S}_-^{(1)}-\hat{S}_+^{(1)}\hat{S}_-^{(2)})/2$ and $\hat{J}_\alpha=\hat{S}_\alpha^{(1)}+\hat{S}_\alpha^{(2)}$ are the total angular momentum operators. Here, $\hat{S}_\alpha^{(1,2)}$ are macroscopic spin operators, each one associated with $N$ two-level systems.

Before analyzing the sensitivity of the second protocol we briefly summarize the most important features of the cascaded dynamical system. A crucial result is that the stationary state of Eq. (\ref{ME_model2}) is a dark state --- for any ratio $\omega/\kappa$ --- as long as $\omega_\mathrm{D}=\omega$ (see  \cite{SM} for its analytical expression). Interestingly, well into the stationary phase, it is separable, while for $\omega/\kappa\gg1$  it tends to the highly correlated singlet state of the total angular momentum. Once we break the condition $\Delta \omega=\omega-\omega_\mathrm{D}=0$, the cascaded system emits light. In Fig.~\ref{fig_protocol_2}(a) we show its stationary emission intensity: $\lim_{T\to\infty} I_T= \kappa\text{Tr}[ \hat{J}_+\hat{J}_-\hat{\rho}_\mathrm{ss}]$. For $\omega_\mathrm{D}/\omega_\mathrm{c}>1$ and $\Delta\omega>0$, the system emits much more intensely than for $\omega_\mathrm{D}/\omega_\mathrm{c}<1$. Looking at the purity of the stationary state [Fig.~\ref{fig_protocol_2}(b)], it is possible to see a clear  separation between these two regions, the brighter one being highly mixed, while the darker one displaying an almost pure stationary state. Notice that the reduced state of the sensor is exactly the same as for Eq.~\eqref{ME_model1} \cite{Stannigel2012}. Hence, the mixed bright phase corresponds to the sensor being in the time-crystal phase, while the pure phase corresponds to the sensor being stationary. Using the HP approximation \cite{SM}, we find that the stationary phase is assumed when $\omega,\omega_{\mathrm{D}}<\omega_\mathrm{c}$ and $\Delta\omega<(\omega_\mathrm{c}-\omega_\mathrm{D})/2$. This result is accurate even for small sizes [see red-dashed line in Fig.~\ref{fig_protocol_2}(a-b)]. In this phase, each collective spin organizes itself in a coherent state with large amplitude plus Gaussian bosonic fluctuations. Their stationary state is a two mode separable state, in which each mode displays analogous properties \cite{SM}.

We proceed by analyzing the sensitivity of Protocol II, which estimates $\omega$ from the output intensity of the cascaded system [see Fig.~\ref{fig_cartoon}]. In Fig.~\ref{fig_protocol_2}(c) we provide the estimation error as function of $\Delta\omega$ for different points of the phase diagram. A first observation is that in the stationary phase there is no improvement with respect to Protocol I. This can be understood again via HP in combination with methods from large deviations (see \cite{SM}) which predicts the estimation error to be: $\overline{\delta \omega} \approx0.5 \sqrt{\kappa}$. The physical interpretation of this result is the same as for the individual system: the emitted light has essentially the same properties as a coherent state.  In contrast, in the cascaded time-crystal phase the sensitivity increases. The estimation error is smaller than in the stationary phase and it improves further with increasing system size, as shown in Fig.~\ref{fig_protocol_2}(c-d). In panel (d) we analyze the system size dependence considering two values of $\Delta \omega$ close to the dark state condition.  By fitting the points for the largest system sizes $N$ (dashed lines), we obtain $\overline{\delta \omega} \propto N^{-\alpha}$. The exponents are $\alpha=0.68$ for $\Delta\omega/\kappa=0.01$ and $\alpha=0.80$ for $\Delta\omega/\kappa=0.005$. The sensitivity thus surpasses the standard quantum limit in the number of particles. For large system sizes the system performs better closer to the dark state point $\Delta\omega=0$. The estimation error displays a non-monotonous behavior with respect to $\Delta \omega$, which reflects the presence of an optimal value for $\overline{\delta\omega}$. This can be understood by analyzing separately the two quantities that contribute to the estimation error [see Fig.~\ref{fig_protocol_2}(e)]:  as system size increases, the derivative of the stationary intensity develops an increasingly sharper peak closer to zero. In contrast, the (time rescaled) standard deviation of the intensity increases monotonously with system size and $\Delta\omega$. The interplay of these two effects makes the optimal estimation error to be found for an intermediate $\Delta\omega$ that depends on $N$.

{\it Conclusions. --} We have investigated parameter estimation through continuously monitoring quantum trajectories of a BTC. We have shown that the time-crystal phase in principle offers enhanced sensitivity, which manifests through the Heisenberg scaling of the QFI ($\propto N^2$)  in the number of particles, $N$. Investigating two protocols, we have shown that a sensitivity surpassing the standard quantum limit can indeed be practically achieved in a cascaded setup of two time-crystals. In view of recent experiments \cite{Ferioli2022}, it would be interesting to consider other continuous monitoring protocols, such as homodyne detection and/or to include finite detection efficiency. Finally, in the presence of local decay,  we expect the observed  phenomena to persist and become more robust  as  the  number  of  atoms increases due to the increasingly large separation of timescales between collective and local processes \cite{SM}.

{\it Acknowledgements. --} We are grateful for financing from the Baden-W\"urttemberg Stiftung through Project No.~BWST\_ISF2019-23. We also acknowledge funding from the Deutsche Forschungsgemeinschaft (DFG, German Research Foundation) under Project No. 435696605 and through the Research Unit FOR 5413/1, Grant No. 465199066. This project has also received funding from the European Union’s Horizon Europe research and innovation program under Grant Agreement No. 101046968 (BRISQ). F.C.~is indebted to the Baden-W\"urttemberg Stiftung for the financial support by the Eliteprogramme for Postdocs. We acknowledge the use of Qutip python library \cite{Qutip1,Qutip2}. We acknowledge support by the state of Baden-Württemberg through bwHPC and the German Research Foundation (DFG) through grant no INST 40/575-1 FUGG (JUSTUS 2 cluster).

\bibliographystyle{apsrev4-2}
\bibliography{references3}

\begin{thebibliography}{81}%
\makeatletter
\providecommand \@ifxundefined [1]{%
 \@ifx{#1\undefined}
}%
\providecommand \@ifnum [1]{%
 \ifnum #1\expandafter \@firstoftwo
 \else \expandafter \@secondoftwo
 \fi
}%
\providecommand \@ifx [1]{%
 \ifx #1\expandafter \@firstoftwo
 \else \expandafter \@secondoftwo
 \fi
}%
\providecommand \natexlab [1]{#1}%
\providecommand \enquote  [1]{``#1''}%
\providecommand \bibnamefont  [1]{#1}%
\providecommand \bibfnamefont [1]{#1}%
\providecommand \citenamefont [1]{#1}%
\providecommand \href@noop [0]{\@secondoftwo}%
\providecommand \href [0]{\begingroup \@sanitize@url \@href}%
\providecommand \@href[1]{\@@startlink{#1}\@@href}%
\providecommand \@@href[1]{\endgroup#1\@@endlink}%
\providecommand \@sanitize@url [0]{\catcode `\\12\catcode `\$12\catcode
  `\&12\catcode `\#12\catcode `\^12\catcode `\_12\catcode `\%12\relax}%
\providecommand \@@startlink[1]{}%
\providecommand \@@endlink[0]{}%
\providecommand \url  [0]{\begingroup\@sanitize@url \@url }%
\providecommand \@url [1]{\endgroup\@href {#1}{\urlprefix }}%
\providecommand \urlprefix  [0]{URL }%
\providecommand \Eprint [0]{\href }%
\providecommand \doibase [0]{https://doi.org/}%
\providecommand \selectlanguage [0]{\@gobble}%
\providecommand \bibinfo  [0]{\@secondoftwo}%
\providecommand \bibfield  [0]{\@secondoftwo}%
\providecommand \translation [1]{[#1]}%
\providecommand \BibitemOpen [0]{}%
\providecommand \bibitemStop [0]{}%
\providecommand \bibitemNoStop [0]{.\EOS\space}%
\providecommand \EOS [0]{\spacefactor3000\relax}%
\providecommand \BibitemShut  [1]{\csname bibitem#1\endcsname}%
\let\auto@bib@innerbib\@empty
%</preamble>
\bibitem [{\citenamefont {Sacha}\ and\ \citenamefont
  {Zakrzewski}(2017)}]{Sacha2018}%
  \BibitemOpen
  \bibfield  {author} {\bibinfo {author} {\bibfnamefont {K.}~\bibnamefont
  {Sacha}}\ and\ \bibinfo {author} {\bibfnamefont {J.}~\bibnamefont
  {Zakrzewski}},\ }\href {https://doi.org/10.1088/1361-6633/aa8b38} {\bibfield
  {journal} {\bibinfo  {journal} {Rep. Prog. Phys.}\ }\textbf {\bibinfo
  {volume} {81}},\ \bibinfo {pages} {016401} (\bibinfo {year}
  {2017})}\BibitemShut {NoStop}%
\bibitem [{\citenamefont {Else}\ \emph {et~al.}(2020)\citenamefont {Else},
  \citenamefont {Monroe}, \citenamefont {Nayak},\ and\ \citenamefont
  {Yao}}]{Else2020}%
  \BibitemOpen
  \bibfield  {author} {\bibinfo {author} {\bibfnamefont {D.~V.}\ \bibnamefont
  {Else}}, \bibinfo {author} {\bibfnamefont {C.}~\bibnamefont {Monroe}},
  \bibinfo {author} {\bibfnamefont {C.}~\bibnamefont {Nayak}},\ and\ \bibinfo
  {author} {\bibfnamefont {N.~Y.}\ \bibnamefont {Yao}},\ }\href
  {https://doi.org/10.1146/annurev-conmatphys-031119-050658} {\bibfield
  {journal} {\bibinfo  {journal} {Annu. Rev. Condens. Matter Phys.}\ }\textbf
  {\bibinfo {volume} {11}},\ \bibinfo {pages} {467} (\bibinfo {year}
  {2020})}\BibitemShut {NoStop}%
\bibitem [{\citenamefont {Zaletel}\ \emph {et~al.}(2023)\citenamefont
  {Zaletel}, \citenamefont {Lukin}, \citenamefont {Monroe}, \citenamefont
  {Nayak}, \citenamefont {Wilczek},\ and\ \citenamefont {Yao}}]{Zaletel2023}%
  \BibitemOpen
  \bibfield  {author} {\bibinfo {author} {\bibfnamefont {M.~P.}\ \bibnamefont
  {Zaletel}}, \bibinfo {author} {\bibfnamefont {M.}~\bibnamefont {Lukin}},
  \bibinfo {author} {\bibfnamefont {C.}~\bibnamefont {Monroe}}, \bibinfo
  {author} {\bibfnamefont {C.}~\bibnamefont {Nayak}}, \bibinfo {author}
  {\bibfnamefont {F.}~\bibnamefont {Wilczek}},\ and\ \bibinfo {author}
  {\bibfnamefont {N.~Y.}\ \bibnamefont {Yao}},\ }\href
  {https://doi.org/10.1103/RevModPhys.95.031001} {\bibfield  {journal}
  {\bibinfo  {journal} {Rev. Mod. Phys.}\ }\textbf {\bibinfo {volume} {95}},\
  \bibinfo {pages} {031001} (\bibinfo {year} {2023})}\BibitemShut {NoStop}%
\bibitem [{\citenamefont {Else}\ \emph {et~al.}(2016)\citenamefont {Else},
  \citenamefont {Bauer},\ and\ \citenamefont {Nayak}}]{Else2016}%
  \BibitemOpen
  \bibfield  {author} {\bibinfo {author} {\bibfnamefont {D.~V.}\ \bibnamefont
  {Else}}, \bibinfo {author} {\bibfnamefont {B.}~\bibnamefont {Bauer}},\ and\
  \bibinfo {author} {\bibfnamefont {C.}~\bibnamefont {Nayak}},\ }\href
  {https://doi.org/10.1103/PhysRevLett.117.090402} {\bibfield  {journal}
  {\bibinfo  {journal} {Phys. Rev. Lett.}\ }\textbf {\bibinfo {volume} {117}},\
  \bibinfo {pages} {090402} (\bibinfo {year} {2016})}\BibitemShut {NoStop}%
\bibitem [{\citenamefont {Khemani}\ \emph {et~al.}(2016)\citenamefont
  {Khemani}, \citenamefont {Lazarides}, \citenamefont {Moessner},\ and\
  \citenamefont {Sondhi}}]{Khemani2016}%
  \BibitemOpen
  \bibfield  {author} {\bibinfo {author} {\bibfnamefont {V.}~\bibnamefont
  {Khemani}}, \bibinfo {author} {\bibfnamefont {A.}~\bibnamefont {Lazarides}},
  \bibinfo {author} {\bibfnamefont {R.}~\bibnamefont {Moessner}},\ and\
  \bibinfo {author} {\bibfnamefont {S.~L.}\ \bibnamefont {Sondhi}},\ }\href
  {https://doi.org/10.1103/PhysRevLett.116.250401} {\bibfield  {journal}
  {\bibinfo  {journal} {Phys. Rev. Lett.}\ }\textbf {\bibinfo {volume} {116}},\
  \bibinfo {pages} {250401} (\bibinfo {year} {2016})}\BibitemShut {NoStop}%
\bibitem [{\citenamefont {Yao}\ \emph {et~al.}(2017)\citenamefont {Yao},
  \citenamefont {Potter}, \citenamefont {Potirniche},\ and\ \citenamefont
  {Vishwanath}}]{Yao2017}%
  \BibitemOpen
  \bibfield  {author} {\bibinfo {author} {\bibfnamefont {N.~Y.}\ \bibnamefont
  {Yao}}, \bibinfo {author} {\bibfnamefont {A.~C.}\ \bibnamefont {Potter}},
  \bibinfo {author} {\bibfnamefont {I.-D.}\ \bibnamefont {Potirniche}},\ and\
  \bibinfo {author} {\bibfnamefont {A.}~\bibnamefont {Vishwanath}},\ }\href
  {https://doi.org/10.1103/PhysRevLett.118.030401} {\bibfield  {journal}
  {\bibinfo  {journal} {Phys. Rev. Lett.}\ }\textbf {\bibinfo {volume} {118}},\
  \bibinfo {pages} {030401} (\bibinfo {year} {2017})}\BibitemShut {NoStop}%
\bibitem [{\citenamefont {Gong}\ \emph {et~al.}(2018)\citenamefont {Gong},
  \citenamefont {Hamazaki},\ and\ \citenamefont {Ueda}}]{Gong2018}%
  \BibitemOpen
  \bibfield  {author} {\bibinfo {author} {\bibfnamefont {Z.}~\bibnamefont
  {Gong}}, \bibinfo {author} {\bibfnamefont {R.}~\bibnamefont {Hamazaki}},\
  and\ \bibinfo {author} {\bibfnamefont {M.}~\bibnamefont {Ueda}},\ }\href
  {https://doi.org/10.1103/PhysRevLett.120.040404} {\bibfield  {journal}
  {\bibinfo  {journal} {Phys. Rev. Lett.}\ }\textbf {\bibinfo {volume} {120}},\
  \bibinfo {pages} {040404} (\bibinfo {year} {2018})}\BibitemShut {NoStop}%
\bibitem [{\citenamefont {Wang}\ \emph {et~al.}(2018)\citenamefont {Wang},
  \citenamefont {Xing}, \citenamefont {Carlo},\ and\ \citenamefont
  {Poletti}}]{Wang2018}%
  \BibitemOpen
  \bibfield  {author} {\bibinfo {author} {\bibfnamefont {R.~R.~W.}\
  \bibnamefont {Wang}}, \bibinfo {author} {\bibfnamefont {B.}~\bibnamefont
  {Xing}}, \bibinfo {author} {\bibfnamefont {G.~G.}\ \bibnamefont {Carlo}},\
  and\ \bibinfo {author} {\bibfnamefont {D.}~\bibnamefont {Poletti}},\ }\href
  {https://doi.org/10.1103/PhysRevE.97.020202} {\bibfield  {journal} {\bibinfo
  {journal} {Phys. Rev. E}\ }\textbf {\bibinfo {volume} {97}},\ \bibinfo
  {pages} {020202} (\bibinfo {year} {2018})}\BibitemShut {NoStop}%
\bibitem [{\citenamefont {Gambetta}\ \emph {et~al.}(2019)\citenamefont
  {Gambetta}, \citenamefont {Carollo}, \citenamefont {Marcuzzi}, \citenamefont
  {Garrahan},\ and\ \citenamefont {Lesanovsky}}]{Gambetta2019}%
  \BibitemOpen
  \bibfield  {author} {\bibinfo {author} {\bibfnamefont {F.~M.}\ \bibnamefont
  {Gambetta}}, \bibinfo {author} {\bibfnamefont {F.}~\bibnamefont {Carollo}},
  \bibinfo {author} {\bibfnamefont {M.}~\bibnamefont {Marcuzzi}}, \bibinfo
  {author} {\bibfnamefont {J.~P.}\ \bibnamefont {Garrahan}},\ and\ \bibinfo
  {author} {\bibfnamefont {I.}~\bibnamefont {Lesanovsky}},\ }\href
  {https://doi.org/10.1103/PhysRevLett.122.015701} {\bibfield  {journal}
  {\bibinfo  {journal} {Phys. Rev. Lett.}\ }\textbf {\bibinfo {volume} {122}},\
  \bibinfo {pages} {015701} (\bibinfo {year} {2019})}\BibitemShut {NoStop}%
\bibitem [{\citenamefont {Lazarides}\ \emph {et~al.}(2020)\citenamefont
  {Lazarides}, \citenamefont {Roy}, \citenamefont {Piazza},\ and\ \citenamefont
  {Moessner}}]{Lazarides2020}%
  \BibitemOpen
  \bibfield  {author} {\bibinfo {author} {\bibfnamefont {A.}~\bibnamefont
  {Lazarides}}, \bibinfo {author} {\bibfnamefont {S.}~\bibnamefont {Roy}},
  \bibinfo {author} {\bibfnamefont {F.}~\bibnamefont {Piazza}},\ and\ \bibinfo
  {author} {\bibfnamefont {R.}~\bibnamefont {Moessner}},\ }\href
  {https://doi.org/10.1103/PhysRevResearch.2.022002} {\bibfield  {journal}
  {\bibinfo  {journal} {Phys. Rev. Research}\ }\textbf {\bibinfo {volume}
  {2}},\ \bibinfo {pages} {022002} (\bibinfo {year} {2020})}\BibitemShut
  {NoStop}%
\bibitem [{\citenamefont {Riera-Campeny}\ \emph {et~al.}(2020)\citenamefont
  {Riera-Campeny}, \citenamefont {Moreno-Cardoner},\ and\ \citenamefont
  {Sanpera}}]{Riera2020}%
  \BibitemOpen
  \bibfield  {author} {\bibinfo {author} {\bibfnamefont {A.}~\bibnamefont
  {Riera-Campeny}}, \bibinfo {author} {\bibfnamefont {M.}~\bibnamefont
  {Moreno-Cardoner}},\ and\ \bibinfo {author} {\bibfnamefont {A.}~\bibnamefont
  {Sanpera}},\ }\href {https://doi.org/10.22331/q-2020-05-25-270} {\bibfield
  {journal} {\bibinfo  {journal} {{Quantum}}\ }\textbf {\bibinfo {volume}
  {4}},\ \bibinfo {pages} {270} (\bibinfo {year} {2020})}\BibitemShut {NoStop}%
\bibitem [{\citenamefont {Zhu}\ \emph {et~al.}(2019)\citenamefont {Zhu},
  \citenamefont {Marino}, \citenamefont {Yao}, \citenamefont {Lukin},\ and\
  \citenamefont {Demler}}]{Zhu2019}%
  \BibitemOpen
  \bibfield  {author} {\bibinfo {author} {\bibfnamefont {B.}~\bibnamefont
  {Zhu}}, \bibinfo {author} {\bibfnamefont {J.}~\bibnamefont {Marino}},
  \bibinfo {author} {\bibfnamefont {N.~Y.}\ \bibnamefont {Yao}}, \bibinfo
  {author} {\bibfnamefont {M.~D.}\ \bibnamefont {Lukin}},\ and\ \bibinfo
  {author} {\bibfnamefont {E.~A.}\ \bibnamefont {Demler}},\ }\href
  {https://doi.org/10.1088/1367-2630/ab2afe} {\bibfield  {journal} {\bibinfo
  {journal} {New J. Phys.}\ }\textbf {\bibinfo {volume} {21}},\ \bibinfo
  {pages} {073028} (\bibinfo {year} {2019})}\BibitemShut {NoStop}%
\bibitem [{\citenamefont {Chinzei}\ and\ \citenamefont
  {Ikeda}(2020)}]{Chinzei2020}%
  \BibitemOpen
  \bibfield  {author} {\bibinfo {author} {\bibfnamefont {K.}~\bibnamefont
  {Chinzei}}\ and\ \bibinfo {author} {\bibfnamefont {T.~N.}\ \bibnamefont
  {Ikeda}},\ }\href {https://doi.org/10.1103/PhysRevLett.125.060601} {\bibfield
   {journal} {\bibinfo  {journal} {Phys. Rev. Lett.}\ }\textbf {\bibinfo
  {volume} {125}},\ \bibinfo {pages} {060601} (\bibinfo {year}
  {2020})}\BibitemShut {NoStop}%
\bibitem [{\citenamefont {Ke\ss{}ler}\ \emph {et~al.}(2021)\citenamefont
  {Ke\ss{}ler}, \citenamefont {Kongkhambut}, \citenamefont {Georges},
  \citenamefont {Mathey}, \citenamefont {Cosme},\ and\ \citenamefont
  {Hemmerich}}]{Kessler2021}%
  \BibitemOpen
  \bibfield  {author} {\bibinfo {author} {\bibfnamefont {H.}~\bibnamefont
  {Ke\ss{}ler}}, \bibinfo {author} {\bibfnamefont {P.}~\bibnamefont
  {Kongkhambut}}, \bibinfo {author} {\bibfnamefont {C.}~\bibnamefont
  {Georges}}, \bibinfo {author} {\bibfnamefont {L.}~\bibnamefont {Mathey}},
  \bibinfo {author} {\bibfnamefont {J.~G.}\ \bibnamefont {Cosme}},\ and\
  \bibinfo {author} {\bibfnamefont {A.}~\bibnamefont {Hemmerich}},\ }\href
  {https://doi.org/10.1103/PhysRevLett.127.043602} {\bibfield  {journal}
  {\bibinfo  {journal} {Phys. Rev. Lett.}\ }\textbf {\bibinfo {volume} {127}},\
  \bibinfo {pages} {043602} (\bibinfo {year} {2021})}\BibitemShut {NoStop}%
\bibitem [{\citenamefont {Tuquero}\ \emph {et~al.}(2022)\citenamefont
  {Tuquero}, \citenamefont {Skulte}, \citenamefont {Mathey},\ and\
  \citenamefont {Cosme}}]{Tuquero2022}%
  \BibitemOpen
  \bibfield  {author} {\bibinfo {author} {\bibfnamefont {R.~J.~L.}\
  \bibnamefont {Tuquero}}, \bibinfo {author} {\bibfnamefont {J.}~\bibnamefont
  {Skulte}}, \bibinfo {author} {\bibfnamefont {L.}~\bibnamefont {Mathey}},\
  and\ \bibinfo {author} {\bibfnamefont {J.~G.}\ \bibnamefont {Cosme}},\ }\href
  {https://doi.org/10.1103/PhysRevA.105.043311} {\bibfield  {journal} {\bibinfo
   {journal} {Phys. Rev. A}\ }\textbf {\bibinfo {volume} {105}},\ \bibinfo
  {pages} {043311} (\bibinfo {year} {2022})}\BibitemShut {NoStop}%
\bibitem [{\citenamefont {Sarkar}\ and\ \citenamefont
  {Dubi}(2022)}]{Sarkar2022}%
  \BibitemOpen
  \bibfield  {author} {\bibinfo {author} {\bibfnamefont {S.}~\bibnamefont
  {Sarkar}}\ and\ \bibinfo {author} {\bibfnamefont {Y.}~\bibnamefont {Dubi}},\
  }\href {https://doi.org/10.1038/s42005-022-00925-z} {\bibfield  {journal}
  {\bibinfo  {journal} {Comm. Phys.}\ }\textbf {\bibinfo {volume} {5}},\
  \bibinfo {pages} {1} (\bibinfo {year} {2022})}\BibitemShut {NoStop}%
\bibitem [{\citenamefont {Cabot}\ \emph {et~al.}(2022)\citenamefont {Cabot},
  \citenamefont {Carollo},\ and\ \citenamefont {Lesanovsky}}]{Cabot2022b}%
  \BibitemOpen
  \bibfield  {author} {\bibinfo {author} {\bibfnamefont {A.}~\bibnamefont
  {Cabot}}, \bibinfo {author} {\bibfnamefont {F.}~\bibnamefont {Carollo}},\
  and\ \bibinfo {author} {\bibfnamefont {I.}~\bibnamefont {Lesanovsky}},\
  }\href {https://doi.org/10.1103/PhysRevB.106.134311} {\bibfield  {journal}
  {\bibinfo  {journal} {Phys. Rev. B}\ }\textbf {\bibinfo {volume} {106}},\
  \bibinfo {pages} {134311} (\bibinfo {year} {2022})}\BibitemShut {NoStop}%
\bibitem [{\citenamefont {Iemini}\ \emph {et~al.}(2018)\citenamefont {Iemini},
  \citenamefont {Russomanno}, \citenamefont {Keeling}, \citenamefont
  {Schir\`o}, \citenamefont {Dalmonte},\ and\ \citenamefont
  {Fazio}}]{Iemini2018}%
  \BibitemOpen
  \bibfield  {author} {\bibinfo {author} {\bibfnamefont {F.}~\bibnamefont
  {Iemini}}, \bibinfo {author} {\bibfnamefont {A.}~\bibnamefont {Russomanno}},
  \bibinfo {author} {\bibfnamefont {J.}~\bibnamefont {Keeling}}, \bibinfo
  {author} {\bibfnamefont {M.}~\bibnamefont {Schir\`o}}, \bibinfo {author}
  {\bibfnamefont {M.}~\bibnamefont {Dalmonte}},\ and\ \bibinfo {author}
  {\bibfnamefont {R.}~\bibnamefont {Fazio}},\ }\href
  {https://doi.org/10.1103/PhysRevLett.121.035301} {\bibfield  {journal}
  {\bibinfo  {journal} {Phys. Rev. Lett.}\ }\textbf {\bibinfo {volume} {121}},\
  \bibinfo {pages} {035301} (\bibinfo {year} {2018})}\BibitemShut {NoStop}%
\bibitem [{\citenamefont {Tucker}\ \emph {et~al.}(2018)\citenamefont {Tucker},
  \citenamefont {Zhu}, \citenamefont {Lewis-Swan}, \citenamefont {Marino},
  \citenamefont {Jimenez}, \citenamefont {Restrepo},\ and\ \citenamefont
  {Rey}}]{Tucker2018}%
  \BibitemOpen
  \bibfield  {author} {\bibinfo {author} {\bibfnamefont {K.}~\bibnamefont
  {Tucker}}, \bibinfo {author} {\bibfnamefont {B.}~\bibnamefont {Zhu}},
  \bibinfo {author} {\bibfnamefont {R.~J.}\ \bibnamefont {Lewis-Swan}},
  \bibinfo {author} {\bibfnamefont {J.}~\bibnamefont {Marino}}, \bibinfo
  {author} {\bibfnamefont {F.}~\bibnamefont {Jimenez}}, \bibinfo {author}
  {\bibfnamefont {J.~G.}\ \bibnamefont {Restrepo}},\ and\ \bibinfo {author}
  {\bibfnamefont {A.~M.}\ \bibnamefont {Rey}},\ }\href
  {https://doi.org/10.1088/1367-2630/aaf18b} {\bibfield  {journal} {\bibinfo
  {journal} {New J. Phys.}\ }\textbf {\bibinfo {volume} {20}},\ \bibinfo
  {pages} {123003} (\bibinfo {year} {2018})}\BibitemShut {NoStop}%
\bibitem [{\citenamefont {Buonaiuto}\ \emph {et~al.}(2021)\citenamefont
  {Buonaiuto}, \citenamefont {Carollo}, \citenamefont {Olmos},\ and\
  \citenamefont {Lesanovsky}}]{Buonaiuto2021}%
  \BibitemOpen
  \bibfield  {author} {\bibinfo {author} {\bibfnamefont {G.}~\bibnamefont
  {Buonaiuto}}, \bibinfo {author} {\bibfnamefont {F.}~\bibnamefont {Carollo}},
  \bibinfo {author} {\bibfnamefont {B.}~\bibnamefont {Olmos}},\ and\ \bibinfo
  {author} {\bibfnamefont {I.}~\bibnamefont {Lesanovsky}},\ }\href
  {https://doi.org/10.1103/PhysRevLett.127.133601} {\bibfield  {journal}
  {\bibinfo  {journal} {Phys. Rev. Lett.}\ }\textbf {\bibinfo {volume} {127}},\
  \bibinfo {pages} {133601} (\bibinfo {year} {2021})}\BibitemShut {NoStop}%
\bibitem [{\citenamefont {Lledó}\ and\ \citenamefont
  {Szymańska}(2020)}]{Lledo2020}%
  \BibitemOpen
  \bibfield  {author} {\bibinfo {author} {\bibfnamefont {C.}~\bibnamefont
  {Lledó}}\ and\ \bibinfo {author} {\bibfnamefont {M.~H.}\ \bibnamefont
  {Szymańska}},\ }\href {https://doi.org/10.1088/1367-2630/ab9ae3} {\bibfield
  {journal} {\bibinfo  {journal} {New J. Phys.}\ }\textbf {\bibinfo {volume}
  {22}},\ \bibinfo {pages} {075002} (\bibinfo {year} {2020})}\BibitemShut
  {NoStop}%
\bibitem [{\citenamefont {Carollo}\ and\ \citenamefont
  {Lesanovsky}(2022)}]{Carollo2022}%
  \BibitemOpen
  \bibfield  {author} {\bibinfo {author} {\bibfnamefont {F.}~\bibnamefont
  {Carollo}}\ and\ \bibinfo {author} {\bibfnamefont {I.}~\bibnamefont
  {Lesanovsky}},\ }\href {https://doi.org/10.1103/PhysRevA.105.L040202}
  {\bibfield  {journal} {\bibinfo  {journal} {Phys. Rev. A}\ }\textbf {\bibinfo
  {volume} {105}},\ \bibinfo {pages} {L040202} (\bibinfo {year}
  {2022})}\BibitemShut {NoStop}%
\bibitem [{\citenamefont {Seibold}\ \emph {et~al.}(2020)\citenamefont
  {Seibold}, \citenamefont {Rota},\ and\ \citenamefont {Savona}}]{Seibold2020}%
  \BibitemOpen
  \bibfield  {author} {\bibinfo {author} {\bibfnamefont {K.}~\bibnamefont
  {Seibold}}, \bibinfo {author} {\bibfnamefont {R.}~\bibnamefont {Rota}},\ and\
  \bibinfo {author} {\bibfnamefont {V.}~\bibnamefont {Savona}},\ }\href
  {https://doi.org/10.1103/PhysRevA.101.033839} {\bibfield  {journal} {\bibinfo
   {journal} {Phys. Rev. A}\ }\textbf {\bibinfo {volume} {101}},\ \bibinfo
  {pages} {033839} (\bibinfo {year} {2020})}\BibitemShut {NoStop}%
\bibitem [{\citenamefont {Bu\ifmmode~\check{c}\else \v{c}\fi{}a}\ and\
  \citenamefont {Jaksch}(2019)}]{Buca2019}%
  \BibitemOpen
  \bibfield  {author} {\bibinfo {author} {\bibfnamefont {B.}~\bibnamefont
  {Bu\ifmmode~\check{c}\else \v{c}\fi{}a}}\ and\ \bibinfo {author}
  {\bibfnamefont {D.}~\bibnamefont {Jaksch}},\ }\href
  {https://doi.org/10.1103/PhysRevLett.123.260401} {\bibfield  {journal}
  {\bibinfo  {journal} {Phys. Rev. Lett.}\ }\textbf {\bibinfo {volume} {123}},\
  \bibinfo {pages} {260401} (\bibinfo {year} {2019})}\BibitemShut {NoStop}%
\bibitem [{\citenamefont {Bu{\v{c}}a}\ \emph {et~al.}(2019)\citenamefont
  {Bu{\v{c}}a}, \citenamefont {Tindall},\ and\ \citenamefont
  {Jaksch}}]{Buca2019b}%
  \BibitemOpen
  \bibfield  {author} {\bibinfo {author} {\bibfnamefont {B.}~\bibnamefont
  {Bu{\v{c}}a}}, \bibinfo {author} {\bibfnamefont {J.}~\bibnamefont
  {Tindall}},\ and\ \bibinfo {author} {\bibfnamefont {D.}~\bibnamefont
  {Jaksch}},\ }\href {https://doi.org/10.1038/s41467-019-09757-y} {\bibfield
  {journal} {\bibinfo  {journal} {Nat. Commun.}\ }\textbf {\bibinfo {volume}
  {10}},\ \bibinfo {pages} {1} (\bibinfo {year} {2019})}\BibitemShut {NoStop}%
\bibitem [{\citenamefont {Booker}\ \emph {et~al.}(2020)\citenamefont {Booker},
  \citenamefont {Bu{\v{c}}a},\ and\ \citenamefont {Jaksch}}]{Booker2020}%
  \BibitemOpen
  \bibfield  {author} {\bibinfo {author} {\bibfnamefont {C.}~\bibnamefont
  {Booker}}, \bibinfo {author} {\bibfnamefont {B.}~\bibnamefont {Bu{\v{c}}a}},\
  and\ \bibinfo {author} {\bibfnamefont {D.}~\bibnamefont {Jaksch}},\ }\href
  {https://doi.org/10.1088/1367-2630/ababc4} {\bibfield  {journal} {\bibinfo
  {journal} {New J. Phys.}\ }\textbf {\bibinfo {volume} {22}},\ \bibinfo
  {pages} {085007} (\bibinfo {year} {2020})}\BibitemShut {NoStop}%
\bibitem [{\citenamefont {Prazeres}\ \emph {et~al.}(2021)\citenamefont
  {Prazeres}, \citenamefont {Souza},\ and\ \citenamefont
  {Iemini}}]{Prazeres2021}%
  \BibitemOpen
  \bibfield  {author} {\bibinfo {author} {\bibfnamefont {L.~F.~d.}\
  \bibnamefont {Prazeres}}, \bibinfo {author} {\bibfnamefont {L.~d.~S.}\
  \bibnamefont {Souza}},\ and\ \bibinfo {author} {\bibfnamefont
  {F.}~\bibnamefont {Iemini}},\ }\href
  {https://doi.org/10.1103/PhysRevB.103.184308} {\bibfield  {journal} {\bibinfo
   {journal} {Phys. Rev. B}\ }\textbf {\bibinfo {volume} {103}},\ \bibinfo
  {pages} {184308} (\bibinfo {year} {2021})}\BibitemShut {NoStop}%
\bibitem [{\citenamefont {Piccitto}\ \emph {et~al.}(2021)\citenamefont
  {Piccitto}, \citenamefont {Wauters}, \citenamefont {Nori},\ and\
  \citenamefont {Shammah}}]{Piccitto2021}%
  \BibitemOpen
  \bibfield  {author} {\bibinfo {author} {\bibfnamefont {G.}~\bibnamefont
  {Piccitto}}, \bibinfo {author} {\bibfnamefont {M.}~\bibnamefont {Wauters}},
  \bibinfo {author} {\bibfnamefont {F.}~\bibnamefont {Nori}},\ and\ \bibinfo
  {author} {\bibfnamefont {N.}~\bibnamefont {Shammah}},\ }\href
  {https://doi.org/10.1103/PhysRevB.104.014307} {\bibfield  {journal} {\bibinfo
   {journal} {Phys. Rev. B}\ }\textbf {\bibinfo {volume} {104}},\ \bibinfo
  {pages} {014307} (\bibinfo {year} {2021})}\BibitemShut {NoStop}%
\bibitem [{\citenamefont {Louren\ifmmode~\mbox{\c{c}}\else \c{c}\fi{}o}\ \emph
  {et~al.}(2022)\citenamefont {Louren\ifmmode~\mbox{\c{c}}\else \c{c}\fi{}o},
  \citenamefont {Prazeres}, \citenamefont {Maciel}, \citenamefont {Iemini},\
  and\ \citenamefont {Duzzioni}}]{Louren2022}%
  \BibitemOpen
  \bibfield  {author} {\bibinfo {author} {\bibfnamefont {A.~C.}\ \bibnamefont
  {Louren\ifmmode~\mbox{\c{c}}\else \c{c}\fi{}o}}, \bibinfo {author}
  {\bibfnamefont {L.~F.~d.}\ \bibnamefont {Prazeres}}, \bibinfo {author}
  {\bibfnamefont {T.~O.}\ \bibnamefont {Maciel}}, \bibinfo {author}
  {\bibfnamefont {F.}~\bibnamefont {Iemini}},\ and\ \bibinfo {author}
  {\bibfnamefont {E.~I.}\ \bibnamefont {Duzzioni}},\ }\href
  {https://doi.org/10.1103/PhysRevB.105.134422} {\bibfield  {journal} {\bibinfo
   {journal} {Phys. Rev. B}\ }\textbf {\bibinfo {volume} {105}},\ \bibinfo
  {pages} {134422} (\bibinfo {year} {2022})}\BibitemShut {NoStop}%
\bibitem [{\citenamefont {Hajdu\ifmmode~\check{s}\else \v{s}\fi{}ek}\ \emph
  {et~al.}(2022)\citenamefont {Hajdu\ifmmode~\check{s}\else \v{s}\fi{}ek},
  \citenamefont {Solanki}, \citenamefont {Fazio},\ and\ \citenamefont
  {Vinjanampathy}}]{Hajdusek2022}%
  \BibitemOpen
  \bibfield  {author} {\bibinfo {author} {\bibfnamefont {M.}~\bibnamefont
  {Hajdu\ifmmode~\check{s}\else \v{s}\fi{}ek}}, \bibinfo {author}
  {\bibfnamefont {P.}~\bibnamefont {Solanki}}, \bibinfo {author} {\bibfnamefont
  {R.}~\bibnamefont {Fazio}},\ and\ \bibinfo {author} {\bibfnamefont
  {S.}~\bibnamefont {Vinjanampathy}},\ }\href
  {https://doi.org/10.1103/PhysRevLett.128.080603} {\bibfield  {journal}
  {\bibinfo  {journal} {Phys. Rev. Lett.}\ }\textbf {\bibinfo {volume} {128}},\
  \bibinfo {pages} {080603} (\bibinfo {year} {2022})}\BibitemShut {NoStop}%
\bibitem [{\citenamefont {Krishna}\ \emph {et~al.}(2023)\citenamefont
  {Krishna}, \citenamefont {Solanki}, \citenamefont
  {Hajdu\ifmmode~\check{s}\else \v{s}\fi{}ek},\ and\ \citenamefont
  {Vinjanampathy}}]{Krishna2023}%
  \BibitemOpen
  \bibfield  {author} {\bibinfo {author} {\bibfnamefont {M.}~\bibnamefont
  {Krishna}}, \bibinfo {author} {\bibfnamefont {P.}~\bibnamefont {Solanki}},
  \bibinfo {author} {\bibfnamefont {M.}~\bibnamefont
  {Hajdu\ifmmode~\check{s}\else \v{s}\fi{}ek}},\ and\ \bibinfo {author}
  {\bibfnamefont {S.}~\bibnamefont {Vinjanampathy}},\ }\href
  {https://doi.org/10.1103/PhysRevLett.130.150401} {\bibfield  {journal}
  {\bibinfo  {journal} {Phys. Rev. Lett.}\ }\textbf {\bibinfo {volume} {130}},\
  \bibinfo {pages} {150401} (\bibinfo {year} {2023})}\BibitemShut {NoStop}%
\bibitem [{\citenamefont {Mattes}\ \emph {et~al.}(2023)\citenamefont {Mattes},
  \citenamefont {Lesanovsky},\ and\ \citenamefont {Carollo}}]{Mattes2023}%
  \BibitemOpen
  \bibfield  {author} {\bibinfo {author} {\bibfnamefont {R.}~\bibnamefont
  {Mattes}}, \bibinfo {author} {\bibfnamefont {I.}~\bibnamefont {Lesanovsky}},\
  and\ \bibinfo {author} {\bibfnamefont {F.}~\bibnamefont {Carollo}},\
  }\href@noop {} {\bibfield  {journal} {\bibinfo  {journal} {arXiv preprint
  arXiv:2303.07725}\ } (\bibinfo {year} {2023})}\BibitemShut {NoStop}%
\bibitem [{\citenamefont {Strogatz}(2018)}]{Strogatz2018}%
  \BibitemOpen
  \bibfield  {author} {\bibinfo {author} {\bibfnamefont {S.~H.}\ \bibnamefont
  {Strogatz}},\ }\href@noop {} {\emph {\bibinfo {title} {Nonlinear dynamics and
  chaos: with applications to physics, biology, chemistry, and engineering}}}\
  (\bibinfo  {publisher} {CRC press},\ \bibinfo {year} {2018})\BibitemShut
  {NoStop}%
\bibitem [{\citenamefont {Kongkhambut}\ \emph {et~al.}(2022)\citenamefont
  {Kongkhambut}, \citenamefont {Skulte}, \citenamefont {Mathey}, \citenamefont
  {Cosme}, \citenamefont {Hemmerich},\ and\ \citenamefont
  {Keßler}}]{Kongkhambut2022}%
  \BibitemOpen
  \bibfield  {author} {\bibinfo {author} {\bibfnamefont {P.}~\bibnamefont
  {Kongkhambut}}, \bibinfo {author} {\bibfnamefont {J.}~\bibnamefont {Skulte}},
  \bibinfo {author} {\bibfnamefont {L.}~\bibnamefont {Mathey}}, \bibinfo
  {author} {\bibfnamefont {J.~G.}\ \bibnamefont {Cosme}}, \bibinfo {author}
  {\bibfnamefont {A.}~\bibnamefont {Hemmerich}},\ and\ \bibinfo {author}
  {\bibfnamefont {H.}~\bibnamefont {Keßler}},\ }\href
  {https://doi.org/10.1126/science.abo3382} {\bibfield  {journal} {\bibinfo
  {journal} {Science}\ }\textbf {\bibinfo {volume} {377}},\ \bibinfo {pages}
  {670} (\bibinfo {year} {2022})}\BibitemShut {NoStop}%
\bibitem [{\citenamefont {Ferioli}\ \emph {et~al.}(2023)\citenamefont
  {Ferioli}, \citenamefont {Glicenstein}, \citenamefont {Ferrier-Barbut},\ and\
  \citenamefont {Browaeys}}]{Ferioli2022}%
  \BibitemOpen
  \bibfield  {author} {\bibinfo {author} {\bibfnamefont {G.}~\bibnamefont
  {Ferioli}}, \bibinfo {author} {\bibfnamefont {A.}~\bibnamefont
  {Glicenstein}}, \bibinfo {author} {\bibfnamefont {I.}~\bibnamefont
  {Ferrier-Barbut}},\ and\ \bibinfo {author} {\bibfnamefont {A.}~\bibnamefont
  {Browaeys}},\ }\href
  {https://doi.org/https://doi.org/10.1038/s41567-023-02064-w} {\bibfield
  {journal} {\bibinfo  {journal} {Nature Physics}\ ,\ \bibinfo {pages} {1}}
  (\bibinfo {year} {2023})}\BibitemShut {NoStop}%
\bibitem [{\citenamefont {Degen}\ \emph {et~al.}(2017)\citenamefont {Degen},
  \citenamefont {Reinhard},\ and\ \citenamefont {Cappellaro}}]{Degen2017}%
  \BibitemOpen
  \bibfield  {author} {\bibinfo {author} {\bibfnamefont {C.~L.}\ \bibnamefont
  {Degen}}, \bibinfo {author} {\bibfnamefont {F.}~\bibnamefont {Reinhard}},\
  and\ \bibinfo {author} {\bibfnamefont {P.}~\bibnamefont {Cappellaro}},\
  }\href {https://doi.org/10.1103/RevModPhys.89.035002} {\bibfield  {journal}
  {\bibinfo  {journal} {Rev. Mod. Phys.}\ }\textbf {\bibinfo {volume} {89}},\
  \bibinfo {pages} {035002} (\bibinfo {year} {2017})}\BibitemShut {NoStop}%
\bibitem [{\citenamefont {Braun}\ \emph {et~al.}(2018)\citenamefont {Braun},
  \citenamefont {Adesso}, \citenamefont {Benatti}, \citenamefont {Floreanini},
  \citenamefont {Marzolino}, \citenamefont {Mitchell},\ and\ \citenamefont
  {Pirandola}}]{Braun2018}%
  \BibitemOpen
  \bibfield  {author} {\bibinfo {author} {\bibfnamefont {D.}~\bibnamefont
  {Braun}}, \bibinfo {author} {\bibfnamefont {G.}~\bibnamefont {Adesso}},
  \bibinfo {author} {\bibfnamefont {F.}~\bibnamefont {Benatti}}, \bibinfo
  {author} {\bibfnamefont {R.}~\bibnamefont {Floreanini}}, \bibinfo {author}
  {\bibfnamefont {U.}~\bibnamefont {Marzolino}}, \bibinfo {author}
  {\bibfnamefont {M.~W.}\ \bibnamefont {Mitchell}},\ and\ \bibinfo {author}
  {\bibfnamefont {S.}~\bibnamefont {Pirandola}},\ }\href
  {https://doi.org/10.1103/RevModPhys.90.035006} {\bibfield  {journal}
  {\bibinfo  {journal} {Rev. Mod. Phys.}\ }\textbf {\bibinfo {volume} {90}},\
  \bibinfo {pages} {035006} (\bibinfo {year} {2018})}\BibitemShut {NoStop}%
\bibitem [{\citenamefont {Ilias}\ \emph {et~al.}(2022)\citenamefont {Ilias},
  \citenamefont {Yang}, \citenamefont {Huelga},\ and\ \citenamefont
  {Plenio}}]{Ilias2022}%
  \BibitemOpen
  \bibfield  {author} {\bibinfo {author} {\bibfnamefont {T.}~\bibnamefont
  {Ilias}}, \bibinfo {author} {\bibfnamefont {D.}~\bibnamefont {Yang}},
  \bibinfo {author} {\bibfnamefont {S.~F.}\ \bibnamefont {Huelga}},\ and\
  \bibinfo {author} {\bibfnamefont {M.~B.}\ \bibnamefont {Plenio}},\ }\href
  {https://doi.org/10.1103/PRXQuantum.3.010354} {\bibfield  {journal} {\bibinfo
   {journal} {PRX Quantum}\ }\textbf {\bibinfo {volume} {3}},\ \bibinfo {pages}
  {010354} (\bibinfo {year} {2022})}\BibitemShut {NoStop}%
\bibitem [{\citenamefont {Macieszczak}\ \emph
  {et~al.}(2016{\natexlab{a}})\citenamefont {Macieszczak}, \citenamefont
  {Gu\ifmmode \mbox{\c{t}}\else \c{t}\fi{}\ifmmode~\u{a}\else \u{a}\fi{}},
  \citenamefont {Lesanovsky},\ and\ \citenamefont
  {Garrahan}}]{Macieszczak2016}%
  \BibitemOpen
  \bibfield  {author} {\bibinfo {author} {\bibfnamefont {K.}~\bibnamefont
  {Macieszczak}}, \bibinfo {author} {\bibfnamefont {M.}~\bibnamefont
  {Gu\ifmmode \mbox{\c{t}}\else \c{t}\fi{}\ifmmode~\u{a}\else \u{a}\fi{}}},
  \bibinfo {author} {\bibfnamefont {I.}~\bibnamefont {Lesanovsky}},\ and\
  \bibinfo {author} {\bibfnamefont {J.~P.}\ \bibnamefont {Garrahan}},\ }\href
  {https://doi.org/10.1103/PhysRevA.93.022103} {\bibfield  {journal} {\bibinfo
  {journal} {Phys. Rev. A}\ }\textbf {\bibinfo {volume} {93}},\ \bibinfo
  {pages} {022103} (\bibinfo {year} {2016}{\natexlab{a}})}\BibitemShut
  {NoStop}%
\bibitem [{\citenamefont {Fern\'andez-Lorenzo}\ and\ \citenamefont
  {Porras}(2017)}]{Fernandez2017}%
  \BibitemOpen
  \bibfield  {author} {\bibinfo {author} {\bibfnamefont {S.}~\bibnamefont
  {Fern\'andez-Lorenzo}}\ and\ \bibinfo {author} {\bibfnamefont
  {D.}~\bibnamefont {Porras}},\ }\href
  {https://doi.org/10.1103/PhysRevA.96.013817} {\bibfield  {journal} {\bibinfo
  {journal} {Phys. Rev. A}\ }\textbf {\bibinfo {volume} {96}},\ \bibinfo
  {pages} {013817} (\bibinfo {year} {2017})}\BibitemShut {NoStop}%
\bibitem [{\citenamefont {Heugel}\ \emph {et~al.}(2019)\citenamefont {Heugel},
  \citenamefont {Biondi}, \citenamefont {Zilberberg},\ and\ \citenamefont
  {Chitra}}]{Heugel2019}%
  \BibitemOpen
  \bibfield  {author} {\bibinfo {author} {\bibfnamefont {T.~L.}\ \bibnamefont
  {Heugel}}, \bibinfo {author} {\bibfnamefont {M.}~\bibnamefont {Biondi}},
  \bibinfo {author} {\bibfnamefont {O.}~\bibnamefont {Zilberberg}},\ and\
  \bibinfo {author} {\bibfnamefont {R.}~\bibnamefont {Chitra}},\ }\href
  {https://doi.org/10.1103/PhysRevLett.123.173601} {\bibfield  {journal}
  {\bibinfo  {journal} {Phys. Rev. Lett.}\ }\textbf {\bibinfo {volume} {123}},\
  \bibinfo {pages} {173601} (\bibinfo {year} {2019})}\BibitemShut {NoStop}%
\bibitem [{\citenamefont {Garbe}\ \emph {et~al.}(2020)\citenamefont {Garbe},
  \citenamefont {Bina}, \citenamefont {Keller}, \citenamefont {Paris},\ and\
  \citenamefont {Felicetti}}]{Garbe2020}%
  \BibitemOpen
  \bibfield  {author} {\bibinfo {author} {\bibfnamefont {L.}~\bibnamefont
  {Garbe}}, \bibinfo {author} {\bibfnamefont {M.}~\bibnamefont {Bina}},
  \bibinfo {author} {\bibfnamefont {A.}~\bibnamefont {Keller}}, \bibinfo
  {author} {\bibfnamefont {M.~G.~A.}\ \bibnamefont {Paris}},\ and\ \bibinfo
  {author} {\bibfnamefont {S.}~\bibnamefont {Felicetti}},\ }\href
  {https://doi.org/10.1103/PhysRevLett.124.120504} {\bibfield  {journal}
  {\bibinfo  {journal} {Phys. Rev. Lett.}\ }\textbf {\bibinfo {volume} {124}},\
  \bibinfo {pages} {120504} (\bibinfo {year} {2020})}\BibitemShut {NoStop}%
\bibitem [{\citenamefont {Di~Candia}\ \emph {et~al.}(2023)\citenamefont
  {Di~Candia}, \citenamefont {Minganti}, \citenamefont {Petrovnin},
  \citenamefont {Paraoanu},\ and\ \citenamefont {Felicetti}}]{Gandia2023}%
  \BibitemOpen
  \bibfield  {author} {\bibinfo {author} {\bibfnamefont {R.}~\bibnamefont
  {Di~Candia}}, \bibinfo {author} {\bibfnamefont {F.}~\bibnamefont {Minganti}},
  \bibinfo {author} {\bibfnamefont {K.}~\bibnamefont {Petrovnin}}, \bibinfo
  {author} {\bibfnamefont {G.}~\bibnamefont {Paraoanu}},\ and\ \bibinfo
  {author} {\bibfnamefont {S.}~\bibnamefont {Felicetti}},\ }\href
  {https://doi.org/https://doi.org/10.1038/s41534-023-00690-z} {\bibfield
  {journal} {\bibinfo  {journal} {npj Quantum Information}\ }\textbf {\bibinfo
  {volume} {9}},\ \bibinfo {pages} {23} (\bibinfo {year} {2023})}\BibitemShut
  {NoStop}%
\bibitem [{\citenamefont {Pavlov}\ \emph {et~al.}(2023)\citenamefont {Pavlov},
  \citenamefont {Porras},\ and\ \citenamefont {Ivanov}}]{Pavlov2023}%
  \BibitemOpen
  \bibfield  {author} {\bibinfo {author} {\bibfnamefont {V.~P.}\ \bibnamefont
  {Pavlov}}, \bibinfo {author} {\bibfnamefont {D.}~\bibnamefont {Porras}},\
  and\ \bibinfo {author} {\bibfnamefont {P.~A.}\ \bibnamefont {Ivanov}},\
  }\href {https://doi.org/10.1088/1402-4896/ace99f} {\bibfield  {journal}
  {\bibinfo  {journal} {Physica Scripta}\ }\textbf {\bibinfo {volume} {98}},\
  \bibinfo {pages} {095103} (\bibinfo {year} {2023})}\BibitemShut {NoStop}%
\bibitem [{\citenamefont {Montenegro}\ \emph {et~al.}(2023)\citenamefont
  {Montenegro}, \citenamefont {Genoni}, \citenamefont {Bayat},\ and\
  \citenamefont {Paris}}]{Montenegro2023}%
  \BibitemOpen
  \bibfield  {author} {\bibinfo {author} {\bibfnamefont {V.}~\bibnamefont
  {Montenegro}}, \bibinfo {author} {\bibfnamefont {M.~G.}\ \bibnamefont
  {Genoni}}, \bibinfo {author} {\bibfnamefont {A.}~\bibnamefont {Bayat}},\ and\
  \bibinfo {author} {\bibfnamefont {M.~G.}\ \bibnamefont {Paris}},\ }\href
  {https://doi.org/10.1038/s42005-023-01423-6} {\bibfield  {journal} {\bibinfo
  {journal} {Communications Physics}\ }\textbf {\bibinfo {volume} {6}},\
  \bibinfo {pages} {304} (\bibinfo {year} {2023})}\BibitemShut {NoStop}%
\bibitem [{\citenamefont {Ilias}\ \emph {et~al.}(2023)\citenamefont {Ilias},
  \citenamefont {Yang}, \citenamefont {Huelga},\ and\ \citenamefont
  {Plenio}}]{Ilias2023}%
  \BibitemOpen
  \bibfield  {author} {\bibinfo {author} {\bibfnamefont {T.}~\bibnamefont
  {Ilias}}, \bibinfo {author} {\bibfnamefont {D.}~\bibnamefont {Yang}},
  \bibinfo {author} {\bibfnamefont {S.~F.}\ \bibnamefont {Huelga}},\ and\
  \bibinfo {author} {\bibfnamefont {M.~B.}\ \bibnamefont {Plenio}},\
  }\href@noop {} {\bibfield  {journal} {\bibinfo  {journal} {arXiv preprint
  arXiv:2304.02050}\ } (\bibinfo {year} {2023})}\BibitemShut {NoStop}%
\bibitem [{\citenamefont {Gambetta}\ and\ \citenamefont
  {Wiseman}(2001)}]{Gambetta2001}%
  \BibitemOpen
  \bibfield  {author} {\bibinfo {author} {\bibfnamefont {J.}~\bibnamefont
  {Gambetta}}\ and\ \bibinfo {author} {\bibfnamefont {H.~M.}\ \bibnamefont
  {Wiseman}},\ }\href {https://doi.org/10.1103/PhysRevA.64.042105} {\bibfield
  {journal} {\bibinfo  {journal} {Phys. Rev. A}\ }\textbf {\bibinfo {volume}
  {64}},\ \bibinfo {pages} {042105} (\bibinfo {year} {2001})}\BibitemShut
  {NoStop}%
\bibitem [{\citenamefont {Kiilerich}\ and\ \citenamefont
  {M\o{}lmer}(2014)}]{Kiilerich2014}%
  \BibitemOpen
  \bibfield  {author} {\bibinfo {author} {\bibfnamefont {A.~H.}\ \bibnamefont
  {Kiilerich}}\ and\ \bibinfo {author} {\bibfnamefont {K.}~\bibnamefont
  {M\o{}lmer}},\ }\href {https://doi.org/10.1103/PhysRevA.89.052110} {\bibfield
   {journal} {\bibinfo  {journal} {Phys. Rev. A}\ }\textbf {\bibinfo {volume}
  {89}},\ \bibinfo {pages} {052110} (\bibinfo {year} {2014})}\BibitemShut
  {NoStop}%
\bibitem [{\citenamefont {Albarelli}\ \emph {et~al.}(2017)\citenamefont
  {Albarelli}, \citenamefont {Rossi}, \citenamefont {Paris},\ and\
  \citenamefont {Genoni}}]{Albarelli2017}%
  \BibitemOpen
  \bibfield  {author} {\bibinfo {author} {\bibfnamefont {F.}~\bibnamefont
  {Albarelli}}, \bibinfo {author} {\bibfnamefont {M.~A.~C.}\ \bibnamefont
  {Rossi}}, \bibinfo {author} {\bibfnamefont {M.~G.~A.}\ \bibnamefont
  {Paris}},\ and\ \bibinfo {author} {\bibfnamefont {M.~G.}\ \bibnamefont
  {Genoni}},\ }\href {https://doi.org/10.1088/1367-2630/aa9840} {\bibfield
  {journal} {\bibinfo  {journal} {New J. Phys.}\ }\textbf {\bibinfo {volume}
  {19}},\ \bibinfo {pages} {123011} (\bibinfo {year} {2017})}\BibitemShut
  {NoStop}%
\bibitem [{\citenamefont {Albarelli}\ \emph {et~al.}(2018)\citenamefont
  {Albarelli}, \citenamefont {Rossi}, \citenamefont {Tamascelli},\ and\
  \citenamefont {Genoni}}]{Albarelli2018}%
  \BibitemOpen
  \bibfield  {author} {\bibinfo {author} {\bibfnamefont {F.}~\bibnamefont
  {Albarelli}}, \bibinfo {author} {\bibfnamefont {M.~A.~C.}\ \bibnamefont
  {Rossi}}, \bibinfo {author} {\bibfnamefont {D.}~\bibnamefont {Tamascelli}},\
  and\ \bibinfo {author} {\bibfnamefont {M.~G.}\ \bibnamefont {Genoni}},\
  }\href {https://doi.org/10.22331/q-2018-12-03-110} {\bibfield  {journal}
  {\bibinfo  {journal} {{Quantum}}\ }\textbf {\bibinfo {volume} {2}},\ \bibinfo
  {pages} {110} (\bibinfo {year} {2018})}\BibitemShut {NoStop}%
\bibitem [{\citenamefont {Shankar}\ \emph {et~al.}(2019)\citenamefont
  {Shankar}, \citenamefont {Greve}, \citenamefont {Wu}, \citenamefont
  {Thompson},\ and\ \citenamefont {Holland}}]{Shankar2019}%
  \BibitemOpen
  \bibfield  {author} {\bibinfo {author} {\bibfnamefont {A.}~\bibnamefont
  {Shankar}}, \bibinfo {author} {\bibfnamefont {G.~P.}\ \bibnamefont {Greve}},
  \bibinfo {author} {\bibfnamefont {B.}~\bibnamefont {Wu}}, \bibinfo {author}
  {\bibfnamefont {J.~K.}\ \bibnamefont {Thompson}},\ and\ \bibinfo {author}
  {\bibfnamefont {M.}~\bibnamefont {Holland}},\ }\href
  {https://doi.org/10.1103/PhysRevLett.122.233602} {\bibfield  {journal}
  {\bibinfo  {journal} {Phys. Rev. Lett.}\ }\textbf {\bibinfo {volume} {122}},\
  \bibinfo {pages} {233602} (\bibinfo {year} {2019})}\BibitemShut {NoStop}%
\bibitem [{\citenamefont {Albarelli}\ \emph {et~al.}(2020)\citenamefont
  {Albarelli}, \citenamefont {Rossi},\ and\ \citenamefont
  {Genoni}}]{Albarelli2020}%
  \BibitemOpen
  \bibfield  {author} {\bibinfo {author} {\bibfnamefont {F.}~\bibnamefont
  {Albarelli}}, \bibinfo {author} {\bibfnamefont {M.~A.~C.}\ \bibnamefont
  {Rossi}},\ and\ \bibinfo {author} {\bibfnamefont {M.~G.}\ \bibnamefont
  {Genoni}},\ }\href {https://doi.org/10.1142/S0219749919410132} {\bibfield
  {journal} {\bibinfo  {journal} {Int. J. Quantum Inf.}\ }\textbf {\bibinfo
  {volume} {18}},\ \bibinfo {pages} {1941013} (\bibinfo {year}
  {2020})}\BibitemShut {NoStop}%
\bibitem [{\citenamefont {Rossi}\ \emph {et~al.}(2020)\citenamefont {Rossi},
  \citenamefont {Albarelli}, \citenamefont {Tamascelli},\ and\ \citenamefont
  {Genoni}}]{Rossi2020}%
  \BibitemOpen
  \bibfield  {author} {\bibinfo {author} {\bibfnamefont {M.~A.~C.}\
  \bibnamefont {Rossi}}, \bibinfo {author} {\bibfnamefont {F.}~\bibnamefont
  {Albarelli}}, \bibinfo {author} {\bibfnamefont {D.}~\bibnamefont
  {Tamascelli}},\ and\ \bibinfo {author} {\bibfnamefont {M.~G.}\ \bibnamefont
  {Genoni}},\ }\href {https://doi.org/10.1103/PhysRevLett.125.200505}
  {\bibfield  {journal} {\bibinfo  {journal} {Phys. Rev. Lett.}\ }\textbf
  {\bibinfo {volume} {125}},\ \bibinfo {pages} {200505} (\bibinfo {year}
  {2020})}\BibitemShut {NoStop}%
\bibitem [{\citenamefont {Nurdin}\ and\ \citenamefont
  {Guţǎ}(2022)}]{Nurdin2022}%
  \BibitemOpen
  \bibfield  {author} {\bibinfo {author} {\bibfnamefont {H.~I.}\ \bibnamefont
  {Nurdin}}\ and\ \bibinfo {author} {\bibfnamefont {M.}~\bibnamefont
  {Guţǎ}},\ }\href
  {https://doi.org/https://doi.org/10.1016/j.arcontrol.2022.04.012} {\bibfield
  {journal} {\bibinfo  {journal} {Annu. Rev. Control}\ }\textbf {\bibinfo
  {volume} {54}},\ \bibinfo {pages} {295} (\bibinfo {year} {2022})}\BibitemShut
  {NoStop}%
\bibitem [{\citenamefont {Gammelmark}\ and\ \citenamefont
  {M\o{}lmer}(2014)}]{Gammelmark2014}%
  \BibitemOpen
  \bibfield  {author} {\bibinfo {author} {\bibfnamefont {S.}~\bibnamefont
  {Gammelmark}}\ and\ \bibinfo {author} {\bibfnamefont {K.}~\bibnamefont
  {M\o{}lmer}},\ }\href {https://doi.org/10.1103/PhysRevLett.112.170401}
  {\bibfield  {journal} {\bibinfo  {journal} {Phys. Rev. Lett.}\ }\textbf
  {\bibinfo {volume} {112}},\ \bibinfo {pages} {170401} (\bibinfo {year}
  {2014})}\BibitemShut {NoStop}%
\bibitem [{\citenamefont {Catana}\ \emph {et~al.}(2015)\citenamefont {Catana},
  \citenamefont {Bouten},\ and\ \citenamefont {Guţă}}]{Catana2015}%
  \BibitemOpen
  \bibfield  {author} {\bibinfo {author} {\bibfnamefont {C.}~\bibnamefont
  {Catana}}, \bibinfo {author} {\bibfnamefont {L.}~\bibnamefont {Bouten}},\
  and\ \bibinfo {author} {\bibfnamefont {M.}~\bibnamefont {Guţă}},\ }\href
  {https://doi.org/10.1088/1751-8113/48/36/365301} {\bibfield  {journal}
  {\bibinfo  {journal} {J. Phys. A Math. Theor.}\ }\textbf {\bibinfo {volume}
  {48}},\ \bibinfo {pages} {365301} (\bibinfo {year} {2015})}\BibitemShut
  {NoStop}%
\bibitem [{\citenamefont {Gammelmark}\ and\ \citenamefont
  {M\o{}lmer}(2013)}]{Gammelmark2013}%
  \BibitemOpen
  \bibfield  {author} {\bibinfo {author} {\bibfnamefont {S.}~\bibnamefont
  {Gammelmark}}\ and\ \bibinfo {author} {\bibfnamefont {K.}~\bibnamefont
  {M\o{}lmer}},\ }\href {https://doi.org/10.1103/PhysRevA.87.032115} {\bibfield
   {journal} {\bibinfo  {journal} {Phys. Rev. A}\ }\textbf {\bibinfo {volume}
  {87}},\ \bibinfo {pages} {032115} (\bibinfo {year} {2013})}\BibitemShut
  {NoStop}%
\bibitem [{\citenamefont {Kiilerich}\ and\ \citenamefont
  {M\o{}lmer}(2016)}]{Kiilerich2016}%
  \BibitemOpen
  \bibfield  {author} {\bibinfo {author} {\bibfnamefont {A.~H.}\ \bibnamefont
  {Kiilerich}}\ and\ \bibinfo {author} {\bibfnamefont {K.}~\bibnamefont
  {M\o{}lmer}},\ }\href {https://doi.org/10.1103/PhysRevA.94.032103} {\bibfield
   {journal} {\bibinfo  {journal} {Phys. Rev. A}\ }\textbf {\bibinfo {volume}
  {94}},\ \bibinfo {pages} {032103} (\bibinfo {year} {2016})}\BibitemShut
  {NoStop}%
\bibitem [{\citenamefont {Fallani}\ \emph {et~al.}(2022)\citenamefont
  {Fallani}, \citenamefont {Rossi}, \citenamefont {Tamascelli},\ and\
  \citenamefont {Genoni}}]{Fallani2022}%
  \BibitemOpen
  \bibfield  {author} {\bibinfo {author} {\bibfnamefont {A.}~\bibnamefont
  {Fallani}}, \bibinfo {author} {\bibfnamefont {M.~A.~C.}\ \bibnamefont
  {Rossi}}, \bibinfo {author} {\bibfnamefont {D.}~\bibnamefont {Tamascelli}},\
  and\ \bibinfo {author} {\bibfnamefont {M.~G.}\ \bibnamefont {Genoni}},\
  }\href {https://doi.org/10.1103/PRXQuantum.3.020310} {\bibfield  {journal}
  {\bibinfo  {journal} {PRX Quantum}\ }\textbf {\bibinfo {volume} {3}},\
  \bibinfo {pages} {020310} (\bibinfo {year} {2022})}\BibitemShut {NoStop}%
\bibitem [{\citenamefont {Yang}\ \emph {et~al.}(2023)\citenamefont {Yang},
  \citenamefont {Huelga},\ and\ \citenamefont {Plenio}}]{Yang2022}%
  \BibitemOpen
  \bibfield  {author} {\bibinfo {author} {\bibfnamefont {D.}~\bibnamefont
  {Yang}}, \bibinfo {author} {\bibfnamefont {S.~F.}\ \bibnamefont {Huelga}},\
  and\ \bibinfo {author} {\bibfnamefont {M.~B.}\ \bibnamefont {Plenio}},\
  }\href {https://doi.org/10.1103/PhysRevX.13.031012} {\bibfield  {journal}
  {\bibinfo  {journal} {Phys. Rev. X}\ }\textbf {\bibinfo {volume} {13}},\
  \bibinfo {pages} {031012} (\bibinfo {year} {2023})}\BibitemShut {NoStop}%
\bibitem [{\citenamefont {Godley}\ and\ \citenamefont
  {Guta}(2023)}]{Godley2023}%
  \BibitemOpen
  \bibfield  {author} {\bibinfo {author} {\bibfnamefont {A.}~\bibnamefont
  {Godley}}\ and\ \bibinfo {author} {\bibfnamefont {M.}~\bibnamefont {Guta}},\
  }\href {https://doi.org/10.22331/q-2023-04-06-973} {\bibfield  {journal}
  {\bibinfo  {journal} {{Quantum}}\ }\textbf {\bibinfo {volume} {7}},\ \bibinfo
  {pages} {973} (\bibinfo {year} {2023})}\BibitemShut {NoStop}%
\bibitem [{\citenamefont {Gardiner}(1993)}]{Gardiner1993}%
  \BibitemOpen
  \bibfield  {author} {\bibinfo {author} {\bibfnamefont {C.~W.}\ \bibnamefont
  {Gardiner}},\ }\href {https://doi.org/10.1103/PhysRevLett.70.2269} {\bibfield
   {journal} {\bibinfo  {journal} {Phys. Rev. Lett.}\ }\textbf {\bibinfo
  {volume} {70}},\ \bibinfo {pages} {2269} (\bibinfo {year}
  {1993})}\BibitemShut {NoStop}%
\bibitem [{\citenamefont {Carmichael}(1993)}]{Carmichael1993}%
  \BibitemOpen
  \bibfield  {author} {\bibinfo {author} {\bibfnamefont {H.~J.}\ \bibnamefont
  {Carmichael}},\ }\href {https://doi.org/10.1103/PhysRevLett.70.2273}
  {\bibfield  {journal} {\bibinfo  {journal} {Phys. Rev. Lett.}\ }\textbf
  {\bibinfo {volume} {70}},\ \bibinfo {pages} {2273} (\bibinfo {year}
  {1993})}\BibitemShut {NoStop}%
\bibitem [{\citenamefont {Stannigel}\ \emph {et~al.}(2012)\citenamefont
  {Stannigel}, \citenamefont {Rabl},\ and\ \citenamefont
  {Zoller}}]{Stannigel2012}%
  \BibitemOpen
  \bibfield  {author} {\bibinfo {author} {\bibfnamefont {K.}~\bibnamefont
  {Stannigel}}, \bibinfo {author} {\bibfnamefont {P.}~\bibnamefont {Rabl}},\
  and\ \bibinfo {author} {\bibfnamefont {P.}~\bibnamefont {Zoller}},\ }\href
  {https://doi.org/10.1088/1367-2630/14/6/063014} {\bibfield  {journal}
  {\bibinfo  {journal} {New J. Phys.}\ }\textbf {\bibinfo {volume} {14}},\
  \bibinfo {pages} {063014} (\bibinfo {year} {2012})}\BibitemShut {NoStop}%
\bibitem [{\citenamefont {Iemini}\ \emph {et~al.}(2023)\citenamefont {Iemini},
  \citenamefont {Fazio},\ and\ \citenamefont {Sanpera}}]{Iemini2023}%
  \BibitemOpen
  \bibfield  {author} {\bibinfo {author} {\bibfnamefont {F.}~\bibnamefont
  {Iemini}}, \bibinfo {author} {\bibfnamefont {R.}~\bibnamefont {Fazio}},\ and\
  \bibinfo {author} {\bibfnamefont {A.}~\bibnamefont {Sanpera}},\ }\href@noop
  {} {\bibfield  {journal} {\bibinfo  {journal} {arXiv preprint
  arXiv:2306.03927}\ } (\bibinfo {year} {2023})}\BibitemShut {NoStop}%
\bibitem [{\citenamefont {Cabot}\ \emph {et~al.}(2023)\citenamefont {Cabot},
  \citenamefont {Muhle}, \citenamefont {Carollo},\ and\ \citenamefont
  {Lesanovsky}}]{Cabot2022}%
  \BibitemOpen
  \bibfield  {author} {\bibinfo {author} {\bibfnamefont {A.}~\bibnamefont
  {Cabot}}, \bibinfo {author} {\bibfnamefont {L.~S.}\ \bibnamefont {Muhle}},
  \bibinfo {author} {\bibfnamefont {F.}~\bibnamefont {Carollo}},\ and\ \bibinfo
  {author} {\bibfnamefont {I.}~\bibnamefont {Lesanovsky}},\ }\href
  {https://doi.org/10.1103/PhysRevA.108.L041303} {\bibfield  {journal}
  {\bibinfo  {journal} {Phys. Rev. A}\ }\textbf {\bibinfo {volume} {108}},\
  \bibinfo {pages} {L041303} (\bibinfo {year} {2023})}\BibitemShut {NoStop}%
\bibitem [{\citenamefont {Agarwal}\ \emph {et~al.}(1977)\citenamefont
  {Agarwal}, \citenamefont {Brown}, \citenamefont {Narducci},\ and\
  \citenamefont {Vetri}}]{Agarwal1977}%
  \BibitemOpen
  \bibfield  {author} {\bibinfo {author} {\bibfnamefont {G.~S.}\ \bibnamefont
  {Agarwal}}, \bibinfo {author} {\bibfnamefont {A.~C.}\ \bibnamefont {Brown}},
  \bibinfo {author} {\bibfnamefont {L.~M.}\ \bibnamefont {Narducci}},\ and\
  \bibinfo {author} {\bibfnamefont {G.}~\bibnamefont {Vetri}},\ }\href
  {https://doi.org/10.1103/PhysRevA.15.1613} {\bibfield  {journal} {\bibinfo
  {journal} {Phys. Rev. A}\ }\textbf {\bibinfo {volume} {15}},\ \bibinfo
  {pages} {1613} (\bibinfo {year} {1977})}\BibitemShut {NoStop}%
\bibitem [{\citenamefont {Narducci}\ \emph {et~al.}(1978)\citenamefont
  {Narducci}, \citenamefont {Feng}, \citenamefont {Gilmore},\ and\
  \citenamefont {Agarwal}}]{Narducci1978}%
  \BibitemOpen
  \bibfield  {author} {\bibinfo {author} {\bibfnamefont {L.~M.}\ \bibnamefont
  {Narducci}}, \bibinfo {author} {\bibfnamefont {D.~H.}\ \bibnamefont {Feng}},
  \bibinfo {author} {\bibfnamefont {R.}~\bibnamefont {Gilmore}},\ and\ \bibinfo
  {author} {\bibfnamefont {G.~S.}\ \bibnamefont {Agarwal}},\ }\href
  {https://doi.org/10.1103/PhysRevA.18.1571} {\bibfield  {journal} {\bibinfo
  {journal} {Phys. Rev. A}\ }\textbf {\bibinfo {volume} {18}},\ \bibinfo
  {pages} {1571} (\bibinfo {year} {1978})}\BibitemShut {NoStop}%
\bibitem [{\citenamefont {Drummond}\ and\ \citenamefont
  {Carmichael}(1978)}]{Drummond1978}%
  \BibitemOpen
  \bibfield  {author} {\bibinfo {author} {\bibfnamefont {P.}~\bibnamefont
  {Drummond}}\ and\ \bibinfo {author} {\bibfnamefont {H.}~\bibnamefont
  {Carmichael}},\ }\href
  {https://doi.org/https://doi.org/10.1016/0030-4018(78)90198-0} {\bibfield
  {journal} {\bibinfo  {journal} {Opt. Commun.}\ }\textbf {\bibinfo {volume}
  {27}},\ \bibinfo {pages} {160} (\bibinfo {year} {1978})}\BibitemShut
  {NoStop}%
\bibitem [{\citenamefont {Carmichael}(1980)}]{Carmichael1980}%
  \BibitemOpen
  \bibfield  {author} {\bibinfo {author} {\bibfnamefont {H.~J.}\ \bibnamefont
  {Carmichael}},\ }\href {https://doi.org/10.1088/0022-3700/13/18/009}
  {\bibfield  {journal} {\bibinfo  {journal} {J. Phys. B At. Mol. Opt.}\
  }\textbf {\bibinfo {volume} {13}},\ \bibinfo {pages} {3551} (\bibinfo {year}
  {1980})}\BibitemShut {NoStop}%
\bibitem [{\citenamefont {Benatti}\ \emph {et~al.}(2018)\citenamefont
  {Benatti}, \citenamefont {Carollo}, \citenamefont {Floreanini},\ and\
  \citenamefont {Narnhofer}}]{benatti2018}%
  \BibitemOpen
  \bibfield  {author} {\bibinfo {author} {\bibfnamefont {F.}~\bibnamefont
  {Benatti}}, \bibinfo {author} {\bibfnamefont {F.}~\bibnamefont {Carollo}},
  \bibinfo {author} {\bibfnamefont {R.}~\bibnamefont {Floreanini}},\ and\
  \bibinfo {author} {\bibfnamefont {H.}~\bibnamefont {Narnhofer}},\ }\href
  {https://doi.org/10.1088/1751-8121/aacbdb} {\bibfield  {journal} {\bibinfo
  {journal} {J. Phys. A: Math. Theor.}\ }\textbf {\bibinfo {volume} {51}},\
  \bibinfo {pages} {325001} (\bibinfo {year} {2018})}\BibitemShut {NoStop}%
\bibitem [{\citenamefont {Wiseman}\ and\ \citenamefont
  {Milburn}(2009)}]{Wiseman2009}%
  \BibitemOpen
  \bibfield  {author} {\bibinfo {author} {\bibfnamefont {H.~M.}\ \bibnamefont
  {Wiseman}}\ and\ \bibinfo {author} {\bibfnamefont {G.~J.}\ \bibnamefont
  {Milburn}},\ }\href@noop {} {\emph {\bibinfo {title} {Quantum measurement and
  control}}}\ (\bibinfo  {publisher} {Cambridge university press},\ \bibinfo
  {year} {2009})\BibitemShut {NoStop}%
\bibitem [{\citenamefont {Garrahan}\ and\ \citenamefont
  {Lesanovsky}(2010)}]{Garrahan2010}%
  \BibitemOpen
  \bibfield  {author} {\bibinfo {author} {\bibfnamefont {J.~P.}\ \bibnamefont
  {Garrahan}}\ and\ \bibinfo {author} {\bibfnamefont {I.}~\bibnamefont
  {Lesanovsky}},\ }\href {https://doi.org/10.1103/PhysRevLett.104.160601}
  {\bibfield  {journal} {\bibinfo  {journal} {Phys. Rev. Lett.}\ }\textbf
  {\bibinfo {volume} {104}},\ \bibinfo {pages} {160601} (\bibinfo {year}
  {2010})}\BibitemShut {NoStop}%
\bibitem [{\citenamefont {Carollo}\ \emph {et~al.}(2018)\citenamefont
  {Carollo}, \citenamefont {Garrahan}, \citenamefont {Lesanovsky},\ and\
  \citenamefont {P\'erez-Espigares}}]{Carollo2018}%
  \BibitemOpen
  \bibfield  {author} {\bibinfo {author} {\bibfnamefont {F.}~\bibnamefont
  {Carollo}}, \bibinfo {author} {\bibfnamefont {J.~P.}\ \bibnamefont
  {Garrahan}}, \bibinfo {author} {\bibfnamefont {I.}~\bibnamefont
  {Lesanovsky}},\ and\ \bibinfo {author} {\bibfnamefont {C.}~\bibnamefont
  {P\'erez-Espigares}},\ }\href {https://doi.org/10.1103/PhysRevA.98.010103}
  {\bibfield  {journal} {\bibinfo  {journal} {Phys. Rev. A}\ }\textbf {\bibinfo
  {volume} {98}},\ \bibinfo {pages} {010103} (\bibinfo {year}
  {2018})}\BibitemShut {NoStop}%
\bibitem [{SM()}]{SM}%
  \BibitemOpen
  \href@noop {} {}\bibinfo {note} {See the Supplemental Material for details,
  which includes Refs. \cite{Kessler2012,Macieszczak2016b}.}\BibitemShut
  {Stop}%
\bibitem [{\citenamefont {Hannukainen}\ and\ \citenamefont
  {Larson}(2018)}]{Hannukainen2018}%
  \BibitemOpen
  \bibfield  {author} {\bibinfo {author} {\bibfnamefont {J.}~\bibnamefont
  {Hannukainen}}\ and\ \bibinfo {author} {\bibfnamefont {J.}~\bibnamefont
  {Larson}},\ }\href {https://doi.org/10.1103/PhysRevA.98.042113} {\bibfield
  {journal} {\bibinfo  {journal} {Phys. Rev. A}\ }\textbf {\bibinfo {volume}
  {98}},\ \bibinfo {pages} {042113} (\bibinfo {year} {2018})}\BibitemShut
  {NoStop}%
\bibitem [{\citenamefont {Passarelli}\ \emph {et~al.}(2023)\citenamefont
  {Passarelli}, \citenamefont {Turkeshi}, \citenamefont {Russomanno},
  \citenamefont {Lucignano}, \citenamefont {Schir{\`o}},\ and\ \citenamefont
  {Fazio}}]{Passarelli2023}%
  \BibitemOpen
  \bibfield  {author} {\bibinfo {author} {\bibfnamefont {G.}~\bibnamefont
  {Passarelli}}, \bibinfo {author} {\bibfnamefont {X.}~\bibnamefont
  {Turkeshi}}, \bibinfo {author} {\bibfnamefont {A.}~\bibnamefont
  {Russomanno}}, \bibinfo {author} {\bibfnamefont {P.}~\bibnamefont
  {Lucignano}}, \bibinfo {author} {\bibfnamefont {M.}~\bibnamefont
  {Schir{\`o}}},\ and\ \bibinfo {author} {\bibfnamefont {R.}~\bibnamefont
  {Fazio}},\ }\href@noop {} {\bibfield  {journal} {\bibinfo  {journal} {arXiv
  preprint arXiv:2306.00841}\ } (\bibinfo {year} {2023})}\BibitemShut {NoStop}%
\bibitem [{\citenamefont {Johansson}\ \emph {et~al.}(2012)\citenamefont
  {Johansson}, \citenamefont {Nation},\ and\ \citenamefont {Nori}}]{Qutip1}%
  \BibitemOpen
  \bibfield  {author} {\bibinfo {author} {\bibfnamefont {J.}~\bibnamefont
  {Johansson}}, \bibinfo {author} {\bibfnamefont {P.}~\bibnamefont {Nation}},\
  and\ \bibinfo {author} {\bibfnamefont {F.}~\bibnamefont {Nori}},\ }\href
  {https://doi.org/https://doi.org/10.1016/j.cpc.2012.02.021} {\bibfield
  {journal} {\bibinfo  {journal} {Comput. Phys. Commun.}\ }\textbf {\bibinfo
  {volume} {183}},\ \bibinfo {pages} {1760} (\bibinfo {year}
  {2012})}\BibitemShut {NoStop}%
\bibitem [{\citenamefont {Johansson}\ \emph {et~al.}(2013)\citenamefont
  {Johansson}, \citenamefont {Nation},\ and\ \citenamefont {Nori}}]{Qutip2}%
  \BibitemOpen
  \bibfield  {author} {\bibinfo {author} {\bibfnamefont {J.}~\bibnamefont
  {Johansson}}, \bibinfo {author} {\bibfnamefont {P.}~\bibnamefont {Nation}},\
  and\ \bibinfo {author} {\bibfnamefont {F.}~\bibnamefont {Nori}},\ }\href
  {https://doi.org/https://doi.org/10.1016/j.cpc.2012.11.019} {\bibfield
  {journal} {\bibinfo  {journal} {Comput. Phys. Commun.}\ }\textbf {\bibinfo
  {volume} {184}},\ \bibinfo {pages} {1234} (\bibinfo {year}
  {2013})}\BibitemShut {NoStop}%
\bibitem [{\citenamefont {Kessler}\ \emph {et~al.}(2012)\citenamefont
  {Kessler}, \citenamefont {Giedke}, \citenamefont {Imamoglu}, \citenamefont
  {Yelin}, \citenamefont {Lukin},\ and\ \citenamefont {Cirac}}]{Kessler2012}%
  \BibitemOpen
  \bibfield  {author} {\bibinfo {author} {\bibfnamefont {E.~M.}\ \bibnamefont
  {Kessler}}, \bibinfo {author} {\bibfnamefont {G.}~\bibnamefont {Giedke}},
  \bibinfo {author} {\bibfnamefont {A.}~\bibnamefont {Imamoglu}}, \bibinfo
  {author} {\bibfnamefont {S.~F.}\ \bibnamefont {Yelin}}, \bibinfo {author}
  {\bibfnamefont {M.~D.}\ \bibnamefont {Lukin}},\ and\ \bibinfo {author}
  {\bibfnamefont {J.~I.}\ \bibnamefont {Cirac}},\ }\href
  {https://doi.org/10.1103/PhysRevA.86.012116} {\bibfield  {journal} {\bibinfo
  {journal} {Phys. Rev. A}\ }\textbf {\bibinfo {volume} {86}},\ \bibinfo
  {pages} {012116} (\bibinfo {year} {2012})}\BibitemShut {NoStop}%
\bibitem [{\citenamefont {Macieszczak}\ \emph
  {et~al.}(2016{\natexlab{b}})\citenamefont {Macieszczak}, \citenamefont
  {Gu\ifmmode \mbox{\c{t}}\else \c{t}\fi{}\ifmmode~\u{a}\else \u{a}\fi{}},
  \citenamefont {Lesanovsky},\ and\ \citenamefont
  {Garrahan}}]{Macieszczak2016b}%
  \BibitemOpen
  \bibfield  {author} {\bibinfo {author} {\bibfnamefont {K.}~\bibnamefont
  {Macieszczak}}, \bibinfo {author} {\bibfnamefont {M.}~\bibnamefont
  {Gu\ifmmode \mbox{\c{t}}\else \c{t}\fi{}\ifmmode~\u{a}\else \u{a}\fi{}}},
  \bibinfo {author} {\bibfnamefont {I.}~\bibnamefont {Lesanovsky}},\ and\
  \bibinfo {author} {\bibfnamefont {J.~P.}\ \bibnamefont {Garrahan}},\ }\href
  {https://doi.org/10.1103/PhysRevLett.116.240404} {\bibfield  {journal}
  {\bibinfo  {journal} {Phys. Rev. Lett.}\ }\textbf {\bibinfo {volume} {116}},\
  \bibinfo {pages} {240404} (\bibinfo {year} {2016}{\natexlab{b}})}\BibitemShut
  {NoStop}%
\end{thebibliography}%

%%%%%%%%%%%%%%%%%%%%%%%%%%%%%%%%%%%%%%%%%%%%%%%%%%%%%%%%%%%%%%%%%%%%%%%%%%%%%%%%%%%%%%%%%%%%%%%%%
%%%%%%%%%%%%%%%%%%%%%%%%%%%%%%%%%%%%%%%%%%%%%%%%%%%%%%%%%%%%%%%%%%%%%%%%%%%%%%%%%%%%%%%%%%%%%%%%%
%%%%%%%%%%%%%%%%%%%%%%%%%%%%%%%%%%%%%%%%%%%%%%%%%%%%%%%%%%%%%%%%%%%%%%%%%%%%%%%%%%%%%%%%%%%%%%%%%
%%%%%%%%%%%%%%%%%%%%%%%%%%%%%%%%%%%%%%%%%%%%%%%%%%%%%%%%%%%%%%%%%%%%%%%%%%%%%%%%%%%%%%%%%%%%%%%%%
%%%%%%%%%%%%%%%%%%%%%%%%%%%%%%%%%%%%%%%%%%%%%%%%%%%%%%%%%%%%%%%%%%%%%%%%%%%%%%%%%%%%%%%%%%%%%%%%%
%%%%%%%%%%%%%%%%%%%%%%%%%%%%%%%%%%%%%%%%%%%%%%%%%%%%%%%%%%%%%%%%%%%%%%%%%%%%%%%%%%%%%%%%%%%%%%%%%
%%%%%%%%%%%%%%%%%%%%%%%%%%%%%%%%%%%%%%%%%%%%%%%%%%%%%%%%%%%%%%%%%%%%%%%%%%%%%%%%%%%%%%%%%%%%%%%%%
%%%%%%%%%%%%%%%%%%%%%%%%%%%%%%%%%%%%%%%%%%%%%%%%%%%%%%%%%%%%%%%%%%%%%%%%%%%%%%%%%%%%%%%%%%%%%%%%%
%%%%%%%%%%%%%%%%%%%%%%%%%%%%%%%%%%%%%%%%%%%%%%%%%%%%%%%%%%%%%%%%%%%%%%%%%%%%%%%%%%%%%%%%%%%%%%%%%
%%%%%%%%%%%%%%%%%%%%%%%%%%%%%%%%%%%%%%%%%%%%%%%%%%%%%%%%%%%%%%%%%%%%%%%%%%%%%%%%%%%%%%%%%%%%%%%%%%%%%%%%%%%%%%%%%%%%%%%%%%%%%%%%%%%%%%%%%%%%%%%%%%%%%%%%%%%%%%%%%%%%%%%%%%%%%%%%%%%%%%%%%%%%%%%%%%
%%%%%%%%%%%%%%%%%%%%%%%%%%%%%%%%%%%%%%%%%%%%%%%%%%%%%%%%%%%%%%%%%%%%%%%%%%%%%%%%%%%%%%%%%%%%%%%%%
%%%%%%%%%%%%%%%%%%%%%%%%%%%%%%%%%%%%%%%%%%%%%%%%%%%%%%%%%%%%%%%%%%%%%%%%%%%%%%%%%%%%%%%%%%%%%%%%%%%%%%%%%%%%%%%%%%%%%%%%%%%%%%%%%%%%%%%%%%%%%%%%%%%%%%%%%%%%%%%%%%%%%%%%%%%%%%%%%%%%%%%%%%%%%%%%%%
%%%%%%%%%%%%%%%%%%%%%%%%%%%%%%%%%%%%%%%%%%%%%%%%%%%%%%%%%%%%%%%%%%%%%%%%%%%%%%%%%%%%%%%%%%%%%%%%%

\setcounter{equation}{0}
\setcounter{figure}{0}
\setcounter{table}{0}
\makeatletter
\renewcommand{\theequation}{S\arabic{equation}}
\renewcommand{\thefigure}{S\arabic{figure}}

\makeatletter
\renewcommand{\theequation}{S\arabic{equation}}
\renewcommand{\thefigure}{S\arabic{figure}}

\onecolumngrid
\newpage

\setcounter{page}{1}

\begin{center}
{\Large SUPPLEMENTAL MATERIAL}
\end{center}
\begin{center}
\vspace{0.8cm}
{\Large Continuous sensing and parameter estimation with the boundary time-crystal}
\end{center}
\begin{center}
Albert Cabot$^{1}$, Federico Carollo$^{1}$, Igor Lesanovsky$^{1,2,3}$
\end{center}
\begin{center}
$^1${\it Institut f\"ur Theoretische Physik, Universit\"at T\"ubingen,}\\
{\it Auf der Morgenstelle 14, 72076 T\"ubingen, Germany}\\
$^2${\it School of Physics and Astronomy, University of Nottingham, Nottingham, NG7 2RD, UK.}\\
$^3${\it Centre for the Mathematics and  Theoretical Physics of Quantum Non-Equilibrium Systems,
University of Nottingham, Nottingham, NG7 2RD, UK}
\end{center}
\section{Calculation of the QFI of the system and emission field}\label{app_fisher}

The QFI of the system and emission field is obtained from the joint system-and-output state. For an arbitrary measurement time $T$ this QFI is given by \cite{Gammelmark2014}:
\begin{equation}
F_\mathrm{E}(g,T)=4\partial_{g_1}\partial_{g_2} \log( |\text{Tr}[\hat{\rho}_{{g_1},{g_2}}(T)]|)\big|_{g_1=g_2=g}. 
\end{equation}
Here $\hat{\rho}_{{g_1},{g_2}}(T)$ is the solution of the following deformed master equation \cite{Gammelmark2014,Macieszczak2016}:
\begin{equation}
\frac{d \hat{\rho}_{{g_1},{g_2}}}{dT}=\mathcal{L}(g_1,g_2)\hat{\rho}_{{g_1},{g_2}}
\end{equation}
with
\begin{equation}\label{deformed_ME}
\mathcal{L}(g_1,g_2)\hat{\rho}=
-i\hat{H}(g_1)\hat{\rho}+i\hat{\rho}\hat{H}(g_2)
+ \hat{L}_j(g_1)\hat{\rho}\hat{L}^\dagger_j(g_2)
-\frac{1}{2}\big( \hat{L}^\dagger_j(g_1)\hat{L}_j(g_1)\hat{\rho}+ \hat{\rho}\hat{L}^\dagger_j(g_2)\hat{L}_j(g_2)\big)
\end{equation}
and initial condition $\hat{\rho}_{{g_1},{g_2}}(0)=\hat{\rho}(0)$. Here we are assuming that the parameter $g$ can be encoded both in the Hamiltonian and jump operators. In this work, we will focus on sensing in the stationary state, i.e. in the long time limit, for which the procedure to obtain the Fisher information simplifies to \cite{Gammelmark2014,Macieszczak2016}:
\begin{equation}\label{fisher_eig}
\lim_{T\to\infty} \frac{F_\mathrm{E}(g,T)}{T}=  4\partial_{g_1}\partial_{g_2} \text{Re}[\lambda_\mathrm{E}(g_1,g_2)]\big|_{g_1=g_2=g}  
\end{equation}
where $\lambda_\mathrm{E}(g_1,g_2)$ is the dominant eigenvalue of $\mathcal{L}(g_1,g_2)$, i.e., the one with the largest real part. As long as the stationary state is unique, for large times the Fisher information of the system and emission field increases linearly. Thus we can see the right hand side of Eq.~\eqref{fisher_eig} as the rate at which this Fisher information increases with time in the asymptotic regime: i.e., the maximum possible rate at which sensitivity can be increased by increasing the measurement time.

\section{Large deviation approach to photocounting}\label{app_largedeviations}

In this work we consider systems described by the standard Markovian master equation in Lindblad form \cite{Wiseman2009}:
\begin{equation}
\partial_t\hat{\rho}=-i[\hat{H},\hat{\rho}]+\mathcal{D}[\hat{L}]\hat{\rho},    
\end{equation}
where $\hat{H}$ is the Hamiltonian of the system and $\hat{L}$ the jump operator. This master equation can be unraveled in a non-linear stochastic Schrödinger equation (SSE), that describes an ideal (unit detection efficiency) photocounting monitoring process \cite{Wiseman2009}:
\begin{equation}
d|\Psi(t)\rangle=dN(t)\bigg(\frac{\hat{L}}{\sqrt{\langle \hat{L}^\dagger \hat{L}\rangle_\mathrm{pc}(t)}}-1\bigg)|\Psi(t)\rangle+[1-dN(t)]dt\bigg(\frac{\langle \hat{L}^\dagger \hat{L}\rangle_\mathrm{pc}(t)}{2}-\frac{\hat{L}^\dagger \hat{L}}{2}-i\hat{H}\bigg)|\Psi(t)\rangle.
\end{equation}
The SSE describes the conditioned (by the measurement protocol) state of the system $|\Psi(t)\rangle$, and expected values with the subscript 'pc' are taken with this state. The quantity $dN(t)$ is a random variable taking the value one if a photon is detected at time $t$ and zero otherwise. Its average value with respect to the instantaneous state in a trajectory is given by:
\begin{equation}
\mathbb{E}[dN(t)]=dt\langle\hat{L}^\dagger\hat{L} \rangle_\mathrm{pc}(t).
\end{equation}

Collecting the number of detections for a measurement time window $T$, we can define the emission or output intensity as
\begin{equation}
I_T=\frac{1}{T}\int_0^T dN(t).
\end{equation}
The time-integrated count record is the central quantity of our sensing protocols. In the long-time measurement limit, it approaches the emission intensity in the stationary state of the master equation:
\begin{equation}\label{long_time_intensity}
\lim_{T\to\infty} \mathbb{E}[I_T] =\text{Tr}[ \hat{L}^\dagger \hat{L} \hat{\rho}_\mathrm{ss}] 
\end{equation}
In the limit $T\to\infty$, the average over realizations in Eq. (\ref{long_time_intensity}) can be dropped out, i.e. the time-integrated quantity tends to its mean, as in a law of large numbers. The way in which this occurs can be understood within the framework of large deviations theory. This is of upmost importance in our sensing approach, as it quantifies how the estimation error changes with measurement time.

In Ref.~\cite{Garrahan2010} a large deviation approach for open quantum systems monitored by ideal photocounting is presented. The central quantity  is the partition function:
\begin{equation}
Z_T(s)=\mathbb{E}\big[e^{-sT I_T}\big], 
\end{equation}
from which we can obtain the moments and cumulants of $I_T$ as partial derivatives with respect to the conjugate field $s$ around $s=0$. Of particular interest for us are the mean and the variance of the emitted intensity:
\begin{equation}
\mathbb{E}[I_T]=-\frac{1}{T}\partial_s Z_T(s)\big|_{s=0},    
\end{equation}
\begin{equation}
\mathbb{E}[I^2_T]-\mathbb{E}[I_T]^2=\frac{1}{T^2}\partial^2_s \log [Z_T(s)]\big|_{s=0}.  
\end{equation}
The partition function can be calculated as $Z_T(s)=\text{Tr}[\hat{\rho}_s(T)]$, where $\hat{\rho}_s(T)$ is the solution of the tilted master equation \cite{Garrahan2010,Carollo2018}:
\begin{equation}\label{tilted_ME}
\frac{d}{dT} \hat{\rho}_s=-i[\hat{H},\hat{\rho}_s]+\mathcal{D}[\hat{L}]\hat{\rho}_s+(e^{-s}-1)\hat{L}\hat{\rho}_s\hat{L}^\dagger. 
\end{equation}
This defines the tilted Liouvilian, $\partial_T\hat{\rho}_s=\mathcal{L}(s)\hat{\rho}_s$. This generator is not trace preserving, though it preserves positivity. Therefore, its dominant eigenvalue, denoted by $\theta(s)$, is real and generally non-zero. This eigenvalue corresponds to the (stationary) {\it scaled cumulant generating function}, and one can obtain all cumulants from its relation with the partition function:
\begin{equation}\label{SCGF}
\lim_{T\to\infty}\frac{1}{T}\log[Z_T(s)]= \theta(s).   
\end{equation}
Thus, for large measurement times $T$:
\begin{equation}
\lim_{T\to\infty}\mathbb{E}[I_T]=-\partial_s \theta(s)|_{s=0},    
\end{equation}
and
\begin{equation}
\lim_{T\to\infty} \mathbb{E}[I^2_T]-\mathbb{E}[I_T]^2=\frac{1}{T}\partial^2_s \theta(s)|_{s=0}.  
\end{equation}
As $\theta(s)$ does not depend on $T$, in the long-time limit the variance diminishes as $T^{-1}$. This means that the estimation error diminishes as $\sqrt{T}$, as stated in the main text. Finally, we find that the prefactor for the  estimation error  can be written entirely in terms of the scaled cummulant generating function:
\begin{equation}
\overline{\delta \omega}=\frac{\sqrt{\theta''(0)}}{|\partial_\omega \theta' (0)|}.    
\end{equation}
This formula combined with a Holstein-Primakoff approach can be used to obtain (approximate) analytical expressions for the estimation error in the stationary phase.

\section{Dark state for the cascaded system}\label{app_darkstate}

In this section we show that the stationary state of the cascaded system is a dark state as long as the sensor and the decoder have the same Rabi frequency, i.e. $\omega=\omega_\mathrm{D}$. For a state to be a dark stationary state, it has to be an eigenstate of the Hamiltonian \cite{Stannigel2012}:
\begin{equation}\label{DS_condition1}
\big[\omega \hat{S}_\mathrm{x}^{(1)}+\omega \hat{S}_\mathrm{x}^{(2)}-i\frac{\kappa}{2}(\hat{S}_+^{(2)}\hat{S}_-^{(1)}-\hat{S}_+^{(1)}\hat{S}_-^{(2)})\big]|\Psi\rangle=\varepsilon_\Psi|\Psi\rangle,    
\end{equation}
and it has to be annihilated by the jump operator:
\begin{equation}\label{DS_condition2}
(\hat{S}_-^{(1)}+\hat{S}_-^{(2)})|\Psi\rangle=0.    
\end{equation}
The second condition implies that the dark state can be written in the basis of the collective angular momentum $\hat{J}_\alpha=\hat{S}_\alpha^{(1)}+\hat{S}_\alpha^{(2)}$ in terms of all states annihilated by $\hat{J}_-$. Hence, the general form for this state reads:
\begin{equation}\label{DS_general}
|\Psi\rangle=\sum_{J=0}^{2S} A_J|J,-J\rangle,  
\end{equation}
where $|J,-J\rangle$ denote the collective angular momentum states with angular momentum $J$ and z-component $-J$, i.e.:
\begin{equation}
|J,-J\rangle=\sum_{m_1,m_2}C^{J,-J}_{S,m_1;S,m_2}|S,m_1;S,m_2\rangle.    
\end{equation}
In the above equation  $C^{J,-J}_{S,m_1;S,m_2}$ denote the Clebsch-Gordan coefficients while $|S,m_1;S,m_2\rangle$ denotes the state in which the z-component of the angular momentum of the sensor is $m_1$ and of the decoder is $m_2$.

{\it General solution.--} The state of Eq. (\ref{DS_general}) satisfies automatically the second condition [Eq. (\ref{DS_condition2})]. The first condition [Eq. (\ref{DS_condition1})] can be satisfied for $\varepsilon_\Psi=0$ if the coefficients $A_\mathrm{J}$ satisfy the following recursive relation:
\begin{eqnarray}\label{DS_recursive}
\omega \sqrt{2J} C^{J,-J+1}_{S,m_1;S,m_2}  A_J=&i\kappa\bigg[
(1-\delta_{-S,m_1+1})(1-\delta_{S,m_2-1})P_+ C^{J-1,-J+1}_{S,m_1+1;S,m_2-1}\\ \nonumber
&-(1-\delta_{S,m_1-1})(1-\delta_{-S,m_2+1})P_- C^{J-1,-J+1}_{S,m_1-1;S,m_2+1}\bigg] A_{J-1}
\end{eqnarray}
with
\small
\begin{equation}
P_\pm=  \sqrt{(S\pm m_1+1)(S\mp m_1)(S\mp m_2+1)(S\pm m_2)}.  
\end{equation}
\normalsize
The previous relation is valid for:
\begin{equation}
J\in[1,2S]\quad \text{and} \quad -J+1=m_1+m_2.    
\end{equation}
Then, the general solution reads:
\begin{equation}\label{DS_solution}
A_{2S}=\frac{1}{\sqrt{\mathcal{N}}}, \quad A_J=\bigg(\frac{\omega}{i\kappa}\bigg)^{2S-J}\frac{a_J}{\sqrt{\mathcal{N}}},  \quad J\in[0,2S-1]. 
\end{equation}
with the normalization constant given by:
\begin{equation}
\mathcal{N}=1+\sum_{J=0}^{2S-1}\bigg(\frac{\omega}{\kappa}\bigg)^{4S-2J} a_J^2.    
\end{equation}
Notice that $a_J$ are numerical constants independent of $\omega$ and $\kappa$, and only determined by $S$ and $J$ through Clebsch-Gordan coefficients and $P_\pm$. 

In the recursive relation [Eq. (\ref{DS_recursive})] the  condition $-J+1=m_1+m_2$ can be satisfied by multiple choices of $m_{1,2}$. All these choices must give an equivalent equation for the $A_J$'s in order for $|\Psi\rangle$ to be an eigenstate of the Hamiltonian. We do not provide an analytical proof for this equivalence. However, we have made a computer program that generates the dark state by solving this recursion relation (for a particular choice of $-J+1=m_1+m_2$) and checks whether it is an eigenstate of the Hamiltonian with zero eigenvalue. This computer program has confirmed so far that this is the case for various (random) choices of $N$ and $\omega/\kappa$. Moreover, we have numerically diagonalized the master equation in the dark state condition $\omega=\omega_\mathrm{D}$ for various points in paramater space, finding that it always displays a unique stationary state. Therefore, we conclude that the dark state obtained from Eqs. (\ref{DS_general}) (\ref{DS_recursive}) and (\ref{DS_solution}) is the stationary state of the system.

{\it Benchmarking with known results. --} As a first check we find that for $S=1/2$, i.e., two spins, $a_0=\sqrt{2}$, and thus:
\begin{equation}
|\Psi\rangle\propto|1,-1\rangle-i\sqrt{2}(\omega/\kappa)|0,0\rangle.    
\end{equation}
This is the same result as found in Ref. \cite{Stannigel2012} (notice a differing factor $2$ due to the fact that they are defining the Hamiltonian with Pauli matrices). A second interesting scenario is that of the limits $\omega/\kappa=0$ and $\omega/\kappa\gg1$. In the former case $|\Psi\rangle=|2S,-2S\rangle$, i.e., all the spins are in the down state. In the latter, we obtain $|\Psi\rangle\approx|0,0\rangle$, which when tracing out the decoder yields the identity matrix for the sensor. Both of these limits coincide with what is known for the individual BTC \cite{Cabot2022}. 

{\it Magnetization of each collective spin. --} The dark state of Eq. (\ref{DS_general}) has some interesting properties with regards to the  individual spin expectation values. In particular, we find that:
\begin{equation}\label{DS_prop1}
\langle \hat{S}_\mathrm{z}^{(j)}\rangle_{\Psi}=\sum_{J=0}^{2S}|A_J|^2 \langle J,-J|  \hat{S}_\mathrm{z}^{(j)}|J,-J\rangle, 
\end{equation}

\begin{equation}\label{DS_prop2}
\langle \hat{S}_-^{(j)}\rangle_{\Psi}=\sum_{J=1}^{2S}A_J^*A_{J-1} \langle J,-J|  \hat{S}_\mathrm{-}^{(j)}|J-1,-J+1\rangle, 
\end{equation}
where $\langle \dots\rangle_\Psi$ denotes an expected value taken with respect to the dark state. The first property [Eq. (\ref{DS_prop1})] comes from the fact that the application of $\hat{S}_\mathrm{z}^{(j)}$ does not change the value $m_1+m_2=-J$ and thus it does not connect different sectors of total z-component. The second property [Eq. (\ref{DS_prop2})] comes from the fact that the application of $\hat{S}_\mathrm{-}^{(j)}$ changes by one the sum $m_1+m_2$ and thus it can only give non-zero contributions when sandwiched by the state of the neighbouring sector  $J+1$.

The states $|J,-J\rangle$ are either symmetric or antisymmetric with respect to the exchange of the labels 1 and 2, i.e. with respect to which is the sensor and which is the decoder. This is because of the following property of the Clebsch-Gordan coefficients:
\begin{equation}
C^{J,M}_{S,m_1;S,m_2}=(-1)^{J-2S} C^{J,M}_{S,m_2;S,m_1}.
\end{equation}
By extension, we find the following two properties:
\begin{equation}\label{CB_prop1}
C^{J,M}_{S,m_1';S,m_2'}C^{J,M}_{S,m_1;S,m_2}=(-1)^{2(J-2S)} C^{J,M}_{S,m_2';S,m_1'}C^{J,M}_{S,m_2;S,m_1}, 
\end{equation}
and 
\begin{equation}\label{CB_prop2}
C^{J,M}_{S,m_1';S,m_2'}C^{J-1,M}_{S,m_1;S,m_2}=
(-1)^{2(J-2S)-1} C^{J,M}_{S,m_2';S,m_1'}C^{J-1,M}_{S,m_2;S,m_1}. 
\end{equation}
From Eq. (\ref{CB_prop1}) it follows that the matrix elements $\langle J,-J|  \hat{S}_\mathrm{z}^{(j)}|J,-J\rangle$ are invariant under exchange of the label $j$. Instead, Eq. (\ref{CB_prop2}) tells us that the matrix elements $\langle J,-J|  \hat{S}_\mathrm{-}^{(j)}|J-1,-J+1\rangle$ change sign under exchange of the label $j$. Therefore, going back to Eqs. (\ref{DS_prop1}) and (\ref{DS_prop2}) we find that:
\begin{equation}
\langle \hat{S}_\mathrm{z}^{(1)}\rangle_{\Psi}=\langle \hat{S}_\mathrm{z}^{(2)}\rangle_{\Psi}, \quad \langle \hat{S}_\mathrm{-}^{(1)}\rangle_{\Psi}=-\langle \hat{S}_\mathrm{-}^{(2)}\rangle_{\Psi}.
\end{equation}
Moreover, since $A_J^* A_{J-1}$ is purely imaginary [see Eq. (\ref{DS_solution})], we obtain:
\begin{equation}
\langle \hat{S}_\mathrm{x}^{(1,2)}\rangle_{\Psi}=0, \quad \langle \hat{S}_\mathrm{y}^{(1)}\rangle_{\Psi}=-\langle \hat{S}_\mathrm{y}^{(2)}\rangle_{\Psi}.    
\end{equation}

\begin{figure}[t!]
 \centering
 \includegraphics[width=0.6\columnwidth]{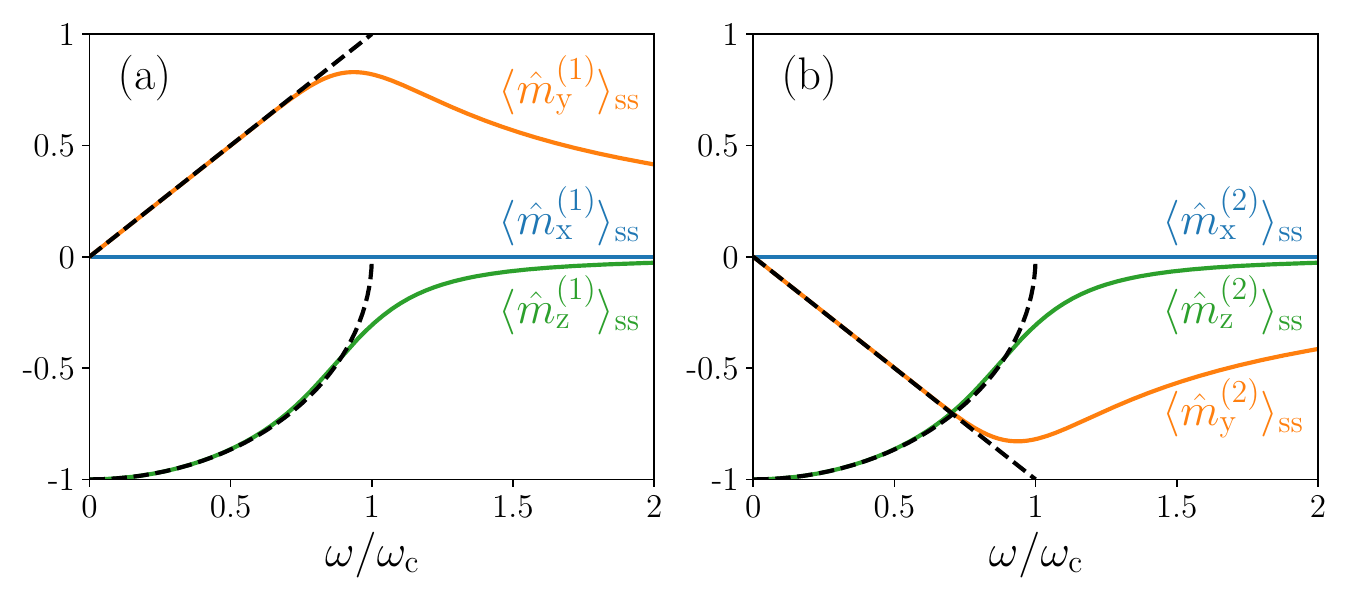}
 \caption{{\bf Stationary state observables for the cascaded system.} Single spin expectation values evaluated in the stationary state of the cascaded system: $\langle \hat{m}_\alpha^{(j)} \rangle_\mathrm{ss}=\langle \hat{S}_\alpha^{(j)} \rangle_\mathrm{ss}/S$. The parameters are fixed to the dark state condition $\omega=\omega_\mathrm{D}$ with $N=10$. The color lines correspond to the exact results solving the master equation. The black broken lines correspond to the mean-field prediction of Eqs. (\ref{DS_MF1}) and (\ref{DS_MF2}).}
 \label{fig_dark_state}
\end{figure}

These results allow us to  apply the mean-field theory for the individual BTC to the cascaded system in the stationary phase. In particular, we know that the reduced state of the sensor is the same of the individual BTC. Thus, the mean-field approximation for $\omega/\omega_\mathrm{c}<1$ yields (see e.g. \cite{Carollo2022}):
\begin{equation}\label{DS_MF1}
\langle \hat{S}_\mathrm{y}^{(1)}\rangle_\mathrm{ss}=\frac{N\omega}{2\omega_\mathrm{c}}, \quad  \langle \hat{S}_\mathrm{z}^{(1)}\rangle_\mathrm{ss}=-\frac{N}{2}\sqrt{1-\frac{\omega^2}{\omega_c^2}}.   
\end{equation}
For the decoder, using the derived properties, we find:
\begin{equation}\label{DS_MF2}
\langle \hat{S}_\mathrm{y}^{(2)}\rangle_\mathrm{ss}=-\frac{N\omega}{2\omega_\mathrm{c}}, \quad  \langle \hat{S}_\mathrm{z}^{(2)}\rangle_\mathrm{ss}=-\frac{N}{2}\sqrt{1-\frac{\omega^2}{\omega_c^2}}.   
\end{equation}
These results are benchmarked in Fig. \ref{fig_dark_state} by solving the master equation numerically. We observe that the y-components invert while the z-ones are identical. We also find that the mean-field results are reasonably accurate given the fact that we are considering small sizes.

\section{Holstein-Primakoff approach for the individual system}\label{app_HP}

The main results of this section are approximate analytical expressions for the estimation error [see Eqs. (\ref{SCGF_individual}) and (\ref{error_analytical_individual})] and the QFI [see Eqs. (\ref{deformed_eig}) and (\ref{QFI_analytical})] in the stationary phase. Moreover, we numerically benchmark these expressions in Figs. \ref{fig_SCGF_check} and \ref{fig_QFI_check}, respectively, finding good agreement. In order to obtain such expressions, we make use of the Holstein-Primakoff (HP) approximation and we apply it to the calculation of  dominant eigenvalue of the tilted master equation [Eq. (\ref{tilted_ME})] and of the deformed master equation [Eq. (\ref{deformed_ME})]. This approach also provides us important physical insights on the interpretation of the results, as discussed in the main text.

\subsection{Holstein-Primakoff approximation}

The starting point of the HP approximation is to express the  spin operators in terms of a bosonic mode (see also Refs. \cite{Kessler2012,Pavlov2023}):
\begin{equation}
\hat{S}_+=\hat{b}^\dagger \sqrt{2S-\hat{b}^\dagger\hat{b}}, \quad  \hat{S}_-= \sqrt{2S-\hat{b}^\dagger\hat{b}}\,\,\hat{b}.
\end{equation}
The bosonic mode is assumed to be in a large displaced state around which we study the fluctuations:
\begin{equation}
\hat{b}\to\hat{b}+\sqrt{S}\beta,    
\end{equation}
where $\beta$ is a complex field to be determined. We proceed to expand the spin operators in powers of the small parameter $\epsilon=1/\sqrt{S}$:
\begin{equation}
\hat{m}_\alpha=\frac{\hat{S}_\alpha}{S}=\sum_{l=0}^\infty \epsilon^l \hat{m}_{\alpha,l}.    
\end{equation}
For our purposes it is only relevant to write down the following explicit expressions:
\begin{equation}
{m}_{+,0}=\sqrt{k}\beta^*,\quad   \hat{m}_{+,1}=\frac{1}{2\sqrt{k}}\big[(4-3|\beta|^2)\hat{b}^\dagger-\beta^{*2}\hat{b} \big],  
\end{equation}
with $k=2-|\beta|^2$.

Before introducing these operators in the master equation and separating by orders, we need to make some physical considerations. First, since the critical point shifts with system size ($\omega_c=\kappa S$), the Rabi frequency can become an extensive parameter, and we need to account for  this explicitly:
\begin{equation}
\omega=\tilde{\omega}S,    
\end{equation}
where $\tilde{\omega}$ is intensive. Due to this fact, the dynamics accelerates with system size and thus time  also needs to be rescaled:
\begin{equation}
\tau=S t.    
\end{equation}
This is essentially the same scaling of parameters prescribed in Ref. \cite{Carmichael1980} for a semiclassical expansion. It is different from the one in Ref. \cite{Pavlov2023} because we consider a fixed decay rate (while there it is scaled with system size).

After the rescaling we arrive to  the following master equation:
\begin{equation}
\partial_\tau\hat{\rho}=-i\tilde{\omega}S[\hat{m}_{x},\hat{\rho}]+\kappa S\big(\hat{m}_-\hat{\rho}\hat{m}_+-\frac{1}{2}\{\hat{m}_+\hat{m}_-,\hat{\rho}\}\big).    
\end{equation}
We now expand the density matrix in powers of $\epsilon$:
\begin{equation}
\hat{\rho}=\sum_{l=0}^\infty    \epsilon^l \hat{\rho}_l
\end{equation}
where Tr$[\hat{\rho}_0]=1$ and Tr$[\hat{\rho}_{l\geq1}]=0$. Then, we plug all the expanded operators and density matrix into the master equation and we separate order by order.

The leading order scales with $\sqrt{S}$ and it provides a self-consistent equation for $\beta$:
\begin{equation}\label{self_consistent}
-i\tilde{\omega}\pm\kappa m_{\pm,0}=0\,\,   \implies\,\, \beta\sqrt{2-|\beta|^2}=i\frac{\tilde{\omega}}{\kappa}.
\end{equation}
Then $m_{\pm,0}=\pm i\tilde{\omega}/\kappa$ and $\beta=-i\sqrt{1-\sqrt{1-(\tilde{\omega}/\kappa)^2}}$. This solution corresponds to the mean-field stable one. Notice that this is only valid in the stationary phase in which $\tilde{\omega}\leq\kappa$.

The next leading term is zero order in $\epsilon$. This describes the leading dynamics of the bosonic fluctuations around the large displaced state. After some algebra one can write this term as  a master equation for the fluctuations:
\begin{equation}\label{fluctuation_ME}
\partial_\tau\hat{\rho}_0=\kappa\big(\hat{m}_{-,1}\hat{\rho}_0\hat{m}_{+,1}-\frac{1}{2}\{\hat{m}_{+,1}\hat{m}_{-,1},\hat{\rho}_0\}\big).    
\end{equation}
For our purposes it is enough to consider this description for the fluctuations and we do not proceed further in the expansion. Notice that this master equation is quadratic, and thus the statistics of the fluctuations are Gaussian. The stationary state is  the vacuum state of the jump operator:
\begin{equation}
\hat{\rho}_{0,\mathrm{ss}}=|E_0\rangle\langle E_0|, \quad \hat{m}_{-,1} |E_0\rangle=0.
\end{equation}
In order to find and show that this state exists in all the stationary phase, we first rewrite more explicitly the jump operator:
\begin{equation}
\hat{m}_{-,1}=A\hat{b}+B\hat{b}^\dagger,    
\end{equation}
with
\small
\begin{equation}
A=\frac{1+3\sqrt{1-(\tilde{\omega}/\kappa)^2}}{2\sqrt{1+\sqrt{1-(\tilde{\omega}/\kappa)^2}}}, \quad B=\frac{1-\sqrt{1-(\tilde{\omega}/\kappa)^2}}{2\sqrt{1+\sqrt{1-(\tilde{\omega}/\kappa)^2}}}.    
\end{equation}
\normalsize
Importantly, $|B/A|\leq 1$. Then, we arrive at the following form for the stationary state:
\begin{equation}\label{fluctuations_ss}
|E_0\rangle=\frac{1}{\sqrt{ \mathcal{N}}}|0\rangle+\frac{1}{\sqrt{ \mathcal{N}}}\sum_{n=1}^\infty(-1)^n\bigg(\frac{B}{A}\bigg)^n\sqrt{\frac{(2n-1)!!}{2n!!}}|2n\rangle,
\end{equation}
where $|n\rangle$ is a fock state and $\mathcal{N}$ is a normalization constant. This state is well defined for $\tilde{\omega}<\kappa$. It actually corresponds to a squeezed state, however, with respect to the emitted light, it appears as a vacuum state as it is annihilated by the jump operator (see also Ref. \cite{Pavlov2023} for an analysis of its properties).

\subsection{Calculation of the scaled cumulant generating function} 

The HP approximation can be combined with large deviations theory in order to obtain approximate expressions for the estimation error bound. We proceed in similar grounds as before but considering the tilted (rescaled) master equation [see Eq. (\ref{tilted_ME})]:
\begin{equation}\label{rescaled_tilted_ME}
\partial_\tau\hat{\rho}=-i\tilde{\omega}S[\hat{m}_{x},\hat{\rho}]+\kappa S\big(\hat{m}_-\hat{\rho}\hat{m}_+-\frac{1}{2}\{\hat{m}_+\hat{m}_-,\hat{\rho}\}\big)
+\kappa S(e^{-s}-1)\hat{m}_-\hat{\rho}\hat{m}_+.
\end{equation}
Our goal is to study this equation around the fixed point we have found previously and in the regime in which the fluctuation master equation is valid [Eq. (\ref{fluctuation_ME})]. This means that we  make use of  the expanded operators we have found in the previous section and we take as the initial condition  the stationary state of the fluctuations, i.e. $\hat{\rho}_{0,\mathrm{ss}}=|E_0\rangle\langle E_0|$. Integrating  Eq. (\ref{rescaled_tilted_ME}) for a small time step $\Delta\tau$  and separating  by orders we obtain:
\begin{equation}
\hat{\rho}(\Delta \tau)\approx\hat{\rho}_{0,\mathrm{ss}}
+\Delta \tau \kappa S(e^{-s}-1)\big[\hat{m}_{-,0}+
+\epsilon\hat{m}_{-,1}+\dots\big]\hat{\rho}_{0,\mathrm{ss}}[\hat{m}_{+,0}+\epsilon\hat{m}_{+,1}+\dots\big].
\end{equation}
Taking into account the properties of $\hat{\rho}_{0,\mathrm{ss}}$ we see that the leading contribution is of order $S$ and that there is no contribution of order $\sqrt{S}$ (as the stationary state is annihilated by the jump operator). Thus, up to order 1 the fluctuation stationary state is an eigenstate of the tilted master equation. From the order $S$ we obtain:
\begin{equation}
\text{Tr}[\hat{\rho}(\tau)]\approx e^{\tilde{\theta}(s)\tau},
\end{equation}
which provides  an approximate expression for the (re-)scaled cumulant generating function
\begin{equation}
\tilde{\theta}(s)=S(e^{-s}-1)\frac{\tilde{\omega}^2}{\kappa}  +\mathcal{O}[1].  
\end{equation}
Taking into account that $\tau=St$, the scaled cumulant generating function in terms of the bare parameters corresponds to:
\begin{equation}
\theta(s)=S \tilde{\theta}(s).    
\end{equation}
which yields:
\begin{equation}\label{SCGF_individual}
{\theta}(s)\approx(e^{-s}-1)\frac{{\omega}^2}{\kappa}.
\end{equation}
From this expression we can obtain the estimation error  presented in the main text:
\begin{equation}\label{error_analytical_individual}
\overline{\delta\omega}=   \frac{\sqrt{\theta''(0)}}{|\partial_\omega \theta' (0)|}\approx 0.5\sqrt{\kappa}.   
\end{equation}
Below we numerically benchmark these expressions finding good agreement.

Finally, we present an alternative approach in order to obtain the (re-)scaled cumulant generating function. This is based on solving the eigenvalue problem for the leading eigenvalue:
\begin{equation}
-i\tilde{\omega}S[\hat{m}_{x},\hat{R}_D]+\kappa S\big(\hat{m}_-\hat{R}_D \hat{m}_+-\frac{1}{2}\{\hat{m}_+\hat{m}_-,\hat{R}_D \}\big)
+\kappa S(e^{-s}-1)\hat{m}_-\hat{R}_D \hat{m}_+ =\tilde{\theta}(s)\hat{R}_D
\end{equation}
in which $\tilde{\theta}(s)$ is the largest eigenvalue and $\hat{R}_D$ the associated eigenvector. The recipe is to expand everything in terms of $\epsilon$ and consider order by order. It is important to notice that the (re-)scaled cumulant generating function can be an extensive parameter, so we expand it as follows:
\begin{equation}
\tilde{\theta}=S\sum_{l=0}^\infty \epsilon^l \tilde{\theta}_l,  \quad \hat{R}_D=\sum_{l=0}^\infty \epsilon^l \hat{R}_{D,l}.     
\end{equation}
where we assume $\text{Tr}[\hat{R}_{D,0}]=1$. At order $S$ we obtain the expression for $l=0$:
\begin{equation}
\tilde{\theta}_0(s)=(e^{-s}-1)\frac{\tilde{\omega}^2}{\kappa}.    
\end{equation}
Here, we have made use of the expression of $\beta$ found before. This can be seen as the contribution from the large displaced state. At order $\sqrt{S}$ the fluctuations kick in  and we obtain the following equation:
\begin{equation}
i\tilde{\omega}(e^{-s}-1)\big(\hat{m}_{-,1}\hat{R}_{D,0}-\hat{R}_{D,0}  \hat{m}_{+,1}\big)=\tilde{\theta}_1  \hat{R}_{D,0}. 
\end{equation}
Hence,  $\hat{R}_{D,0}$ must be an eigenstate of the jump operator: $\hat{m}_{-,1} |E_j\rangle=E_j|E_j\rangle$. Therefore:
\begin{equation}
-2\tilde{\omega}(e^{-s}-1)\text{Im}[E_j]|E_j\rangle \langle E_j|=\tilde{\theta}_1  |E_j\rangle \langle E_j|. 
\end{equation}
This is the general eigenvalue problem at first order. However, we are only interested in one of the solutions. Below we show that $\tilde{\theta}_1'(s=0)=0$.  Given the fact that the eigenvalues $E_j$ do not depend on $s$ (the jump operator is independent of $s$),  the solution corresponding to the (re-)scaled cumulant generating function for the fluctuations is:
\begin{equation}
\tilde{\theta}_1(s)=0,    
\end{equation}
and $\hat{R}_{D,0}=|E_0\rangle \langle E_0|$. 

In order to show that $\tilde{\theta}_1'(s=0)=0$, we recall that the stationary state  allows us to compute this quantity: 
\begin{equation}
\kappa \text{Tr}[ \hat{S}_+\hat{S}_-\hat{\rho}_\mathrm{ss}]=-\theta'(s=0). 
\end{equation}
Expressing it as $S\kappa \text{Tr}[\hat{m}_+\hat{m}_- \hat{\rho}_\mathrm{ss}]=-\tilde{\theta}'(s=0)$ and expanding in orders of $S$ yields:
\begin{equation}
S\kappa m_{+,0}m_{-,0}=S\tilde{\theta}_0'(s=0)
\end{equation}
\begin{equation}
\sqrt{S}\kappa\text{Tr}[\hat{m}_{+,1}{m}_{-,0}\hat{\rho}_{0,\mathrm{ss}}+{m}_{+,0}\hat{m}_{-,1}\hat{\rho}_{0,\mathrm{ss}}]= \sqrt{S}\tilde{\theta}_1'(s=0),
\end{equation}
plus higher orders that are not relevant. Taking into account that $\hat{\rho}_{0,\mathrm{ss}}$ is the vacuum state of the jump operator, we obtain that $\tilde{\theta}_1'(s=0)=0$. Therefore, to leading order in $S$, the eigenvalue method gives:
\begin{equation}
\tilde{\theta}(s)=S(e^{-s}-1)\frac{\tilde{\omega}^2}{\kappa}  +\mathcal{O}[1],  
\end{equation}
which is consistent with the method based on following the dynamics that we have used before.

\begin{figure}[t!]
 \centering
 \includegraphics[width=0.6\columnwidth]{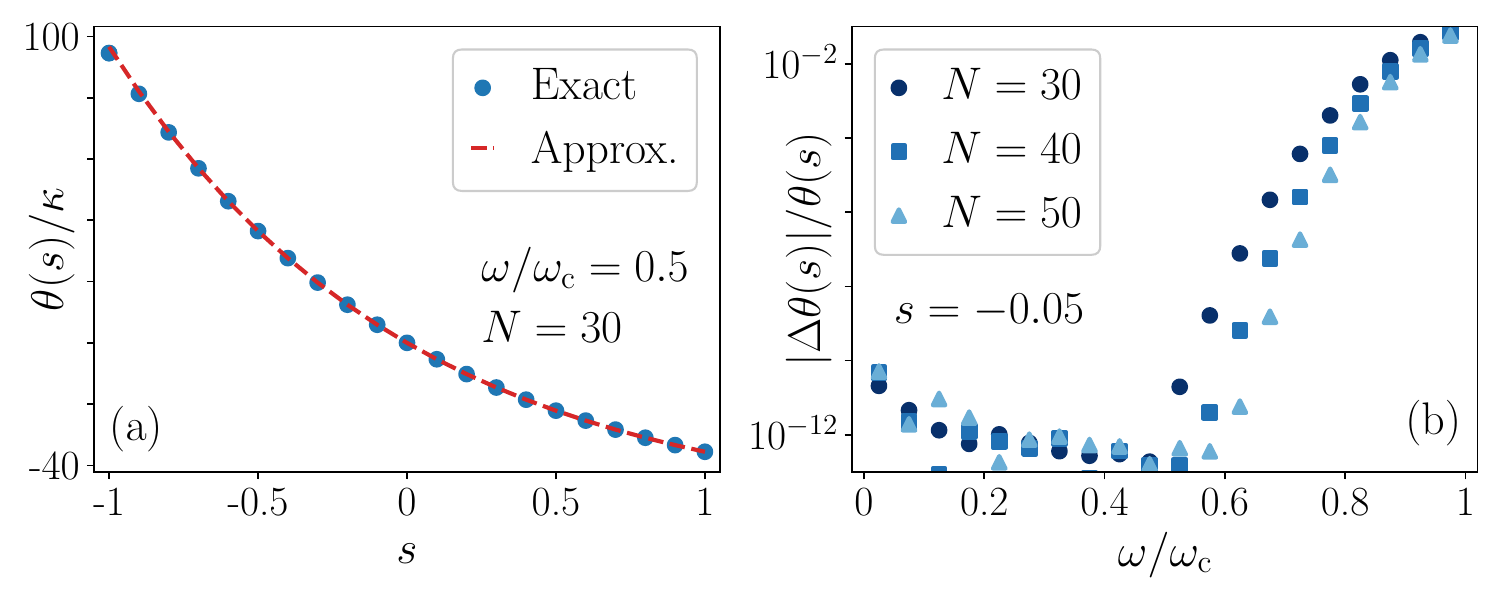}
 \caption{{\bf Benchmark approximate expression for the scaled cumulant generating function.} (a) Scaled cumulant generating function [Eq. (\ref{SCGF})] varying $s$ for $\omega/\omega_\mathrm{c}=0.5$ and $N=30$. The blue points are obtained diagonalizing the tilted master equation. The red dashed line corresponds to the analytical result of Eq. (\ref{SCGF_individual}). (b) Relative error on $\theta(s)$ as a function of $\omega/\omega_\mathrm{c}$, for different system sizes and $s=-0.05$. This error is defined as the difference between the exact result and the analytical result of  Eq. (\ref{SCGF_individual}), both divided by the exact result.  We observe that the error increases near the transition, however, it is generally much smaller than $1/N$.}
 \label{fig_SCGF_check}
\end{figure}

{\it Numerical benchmark. --}   In Fig. \ref{fig_SCGF_check} we compare Eq. (\ref{SCGF_individual}) with the exact result obtained by diagonalizing the corresponding tilted master equation. In panel (a) we show it as a function of $s$ for a fixed $N$ and $\omega/\omega_\mathrm{c}$. In panel (b) we study the relative error, relative difference between exact and approximation, for different system sizes and along the stationary phase. We observe the relative error to be very small in most of the phase diagram, while it increases near the transition. The larger is the system size, the more accurate is our approximation when going closer to the phase transition. Interestingly, the order of the error in Fig. \ref{fig_SCGF_check} (b) suggests that  our approximate results hold even for higher orders in the (system size) perturbative expansion, being virtually exact for small $\omega/\omega_\mathrm{c}$. This can be interpreted as the collective spin system being able to imitate perfectly a coherent mode when not too close from saturation.

\subsection{Calculation of the QFI}

Proceeding analogously we can obtain an approximate expression for the QFI in the stationary phase. In this case we compute the dominant eigenvalue of the deformed master equation for our parameter estimation problem [see Eq. (\ref{deformed_ME})]:
\begin{equation}\label{deformed_ME_spin}
\begin{split}
\partial_t \hat{\rho}=&
-i\omega_1\hat{S}_\mathrm{x}\hat{\rho}+i\omega_2\hat{\rho}\hat{S}_\mathrm{x}
+ \kappa\big(\hat{S}_-\hat{\rho}\hat{S}_+-\frac{1}{2}\{\hat{S}_+\hat{S}_-,\hat{\rho} \}\big).  
\end{split}
\end{equation}
As before, we make use of the rescaled parameters:
\begin{equation}\label{deformed_ME_spin_rescaled}
\partial_\tau \hat{\rho}=
-i\tilde{\omega}_1 S\hat{m}_\mathrm{x}\hat{\rho}+i\tilde{\omega}_2 S\hat{\rho}\hat{m}_\mathrm{x}
+ \kappa S\big(\hat{m}_-\hat{\rho}\hat{m}_+-\frac{1}{2}\{\hat{m}_+\hat{m}_-,\hat{\rho} \}\big).
\end{equation}
We make the following ansatz for the dominant eigenstate of the deformed master equation:
\begin{equation}\label{ansatz_QFI}
\hat{\rho}_0=|E_0^{\tilde{\omega}_1}\rangle\langle E_0^{\tilde{\omega}_2}|.    
\end{equation}
We use $|E_0^{\tilde{\omega}_j}\rangle$ to denote the stationary state of the fluctuations [Eq. (\ref{fluctuations_ss})] at the phase diagram point $\tilde{\omega}_j$. For the condition $\tilde{\omega}_1=\tilde{\omega}_2=\tilde{\omega}$ this state is the stationary state of the fluctuations. We are interested in obtaining an expression around this point, as the QFI corresponds to the curvature of this eigenvalue around it [see Eq. (\ref{fisher_eig})].

When applying the approximated spin operators to this state  we must be careful, as the Bra and the Ket in Eq. (\ref{ansatz_QFI}) describe the fluctuations at different points of the phase diagram i.e. at $\tilde{\omega}_1$ or $\tilde{\omega}_2$. This implies that, e.g.:
\begin{equation}
\hat{m}_-|E_0^{\tilde{\omega}_1}\rangle=\big[ {m}_{-,0}^{\tilde{\omega}_1}+\frac{1}{\sqrt{S}} \hat{m}_{-,1}^{\tilde{\omega}_1}+\dots\big] |E_0^{\tilde{\omega}_1}\rangle, 
\end{equation}
while
\begin{equation}
\langle E_0^{\tilde{\omega}_2}| \hat{m}_-= \langle E_0^{\tilde{\omega}_2}|\big[ {m}_{-,0}^{\tilde{\omega}_2}+\frac{1}{\sqrt{S}} \hat{m}_{-,1}^{\tilde{\omega}_2}+\dots\big], 
\end{equation}
where $\hat{m}_{-,1}^{\tilde{\omega}_j}$ denote the HP approximated operators with a $\beta$ determined for the point in parameter space with $\tilde{\omega}=\tilde{\omega}_j$. Taking into account this, we can introduce our ansatz into the deformed master equation (\ref{deformed_ME_spin_rescaled}) and expand order by order. Then, the right hand side of Eq. (\ref{deformed_ME_spin_rescaled}) gives the following terms. The dominant term:
\begin{equation}
S\bigg[ \frac{\tilde{\omega}_1\tilde{\omega}_2}{\kappa}-\frac{\tilde{\omega}_1^2+\tilde{\omega}_2^2}{2\kappa}\bigg]  |E_0^{\tilde{\omega}_1}\rangle\langle E_0^{\tilde{\omega}_2}|.  
\end{equation}
The subdominant term:
\begin{eqnarray}
\frac{\sqrt{S}}{2}\bigg[-i(\tilde{\omega}_1+\kappa{m}_{-,0}^{\tilde{\omega}_1}) \hat{m}_{+,1}^{\tilde{\omega}_1}
-i(\tilde{\omega}_1-\kappa{m}_{+,0}^{\tilde{\omega}_1})\hat{m}_{-,1}^{\tilde{\omega}_1}\bigg]
|E_0^{\tilde{\omega}_1}\rangle\langle E_0^{\tilde{\omega}_2}|\\ \nonumber
+\frac{\sqrt{S}}{2}|E_0^{\tilde{\omega}_1}\rangle\langle E_0^{\tilde{\omega}_2}| \bigg[i(\tilde{\omega}_2+\kappa{m}_{-,0}^{\tilde{\omega}_2}) \hat{m}_{+,1}^{\tilde{\omega}_2}
+i(\tilde{\omega}_1-\kappa{m}_{+,0}^{\tilde{\omega}_2})\hat{m}_{-,1}^{\tilde{\omega}_2}\bigg]. 
\end{eqnarray}
This term is zero because of the self-consistency condition [Eq. (\ref{self_consistent})]. We do not go beyond this order. Setting the eigenvalue problem in an analogous way as for the scaled cumulant generating function, we obtain that the eigenvalue associated to $|E_0^{\tilde{\omega}_1}\rangle\langle E_0^{\tilde{\omega}_2}|$ is:
\begin{equation}
\tilde{\lambda}_\mathrm{0}(\tilde{\omega}_1,\tilde{\omega}_2)=S\bigg[ \frac{\tilde{\omega}_1\tilde{\omega}_2}{\kappa}-\frac{\tilde{\omega}_1^2+\tilde{\omega}_2^2}{2\kappa}\bigg]+\mathcal{O}[1].    
\end{equation}
We now argue that since $|E_0^{\tilde{\omega}_1}\rangle\langle E_0^{\tilde{\omega}_2}|$ corresponds to the stationary state when $\tilde{\omega}_1=\tilde{\omega}_2$, then very close to this point it must correspond to the dominant eigenmode of the deformed master equation [Eq. (\ref{deformed_ME_spin_rescaled})]  (if the eigenspectrum is gapped, as it is the case in the stationary phase). Based on this argument we make the association:
\begin{equation}
\tilde{\lambda}_\mathrm{E}  (\tilde{\omega}_1,\tilde{\omega}_2) =\tilde{\lambda}_\mathrm{0}(\tilde{\omega}_1,\tilde{\omega}_2), 
\end{equation}
which allows us to obtain the QFI through Eq. (\ref{fisher_eig}). Before that, however, we must reverse the scaling of time ($\tau=St)$, i.e.:
\begin{equation}
{\lambda}_\mathrm{E}  ({\omega}_1,{\omega}_2) =S\tilde{\lambda}_\mathrm{E}(\tilde{\omega}_1,\tilde{\omega}_2), 
\end{equation}
obtaining:
\begin{equation}\label{deformed_eig}
\lambda_\mathrm{E}(\omega_1,\omega_2)\approx\frac{\omega_1\omega_2}{\kappa}-\frac{\omega_1^2+\omega_2^2}{2\kappa}.    
\end{equation}
Then, the QFI in the stationary phase and for long times reads [see Eq. (\ref{fisher_eig})]:
\begin{equation}\label{QFI_analytical}
F_\mathrm{E}(\omega,t)\approx\frac{4t}{\kappa},    
\end{equation}
as stated in the main text.

\begin{figure}[t!]
 \centering
 \includegraphics[width=0.6\columnwidth]{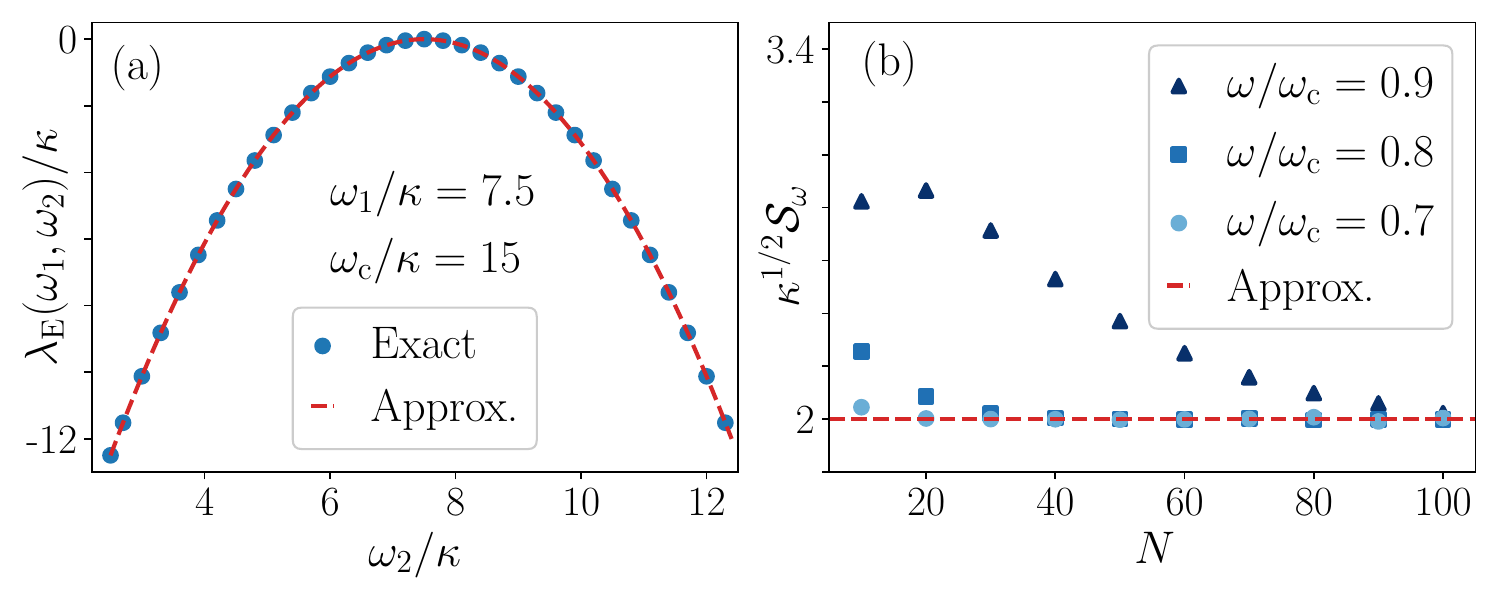}
 \caption{{\bf Benchmark approximate expression for the QFI.} (a) Dominant eigenvalue of the deformed master equation [Eq. (\ref{deformed_ME_spin})] as a function of $\omega_2$ and fixing $\omega_1/\kappa=7.5$ (i.e. $\omega_1/\omega_\mathrm{c}=0.5$) and $N=30$. The blue points are obtained diagonalizing the deformed generator. The red dashed line corresponds to the analytical result of Eq. (\ref{deformed_eig}). (b) Sensitivity bound  $\mathcal{S}_\omega$ [Eq. (\ref{sensitivity_bound})] as a function of system size. The points correspond to different values of $\omega/\omega_\mathrm{c}$ close to the bifurcation. The red dashed line corresponds to the prediction derived from Eq. (\ref{QFI_analytical}). We observe that the closer to the phase transition we are, the larger needs to be $N$ in order for the HP approximation to be accurate.}
 \label{fig_QFI_check}
\end{figure}

{\it Numerical benchmark. --} We numerically benchmark Eq. (\ref{deformed_eig})  in Fig. \ref{fig_QFI_check} (a). We observe a good agreement even far from the the condition $\omega_1=\omega_2$. Moreover, by analyzing the relative error we observe again that this is very small, suggesting that our approximations (far from the critical point) hold even for higher orders in the (system size) perturbative expansion (not shown). In Fig. \ref{fig_QFI_check} (b) we focus on the sensitivity bound, as derived from Eq. (\ref{QFI_analytical}). In this panel we illustrate that the closer one gets to the transition, the larger needs to be the system size in order for the analytical approximate results to be valid.

\section{Holstein-Primakoff approach for the  Cascaded system}\label{app_HP_c}

In this section we apply the Holstein-Primakoff to the cascaded system in order to understand its stationary phase. The main result of this section is an approximate analytical expression for the estimation error [see Eqs.~\eqref{SCGF_cascaded} and \eqref{error_analytical_cascaded}] in the stationary phase. We numerically benchmark the analytical results in Fig.~\ref{fig_SCGF_cascaded}, finding good agreement.

\subsection{Holstein-Primakoff approximation}

We proceed in a similar way as in the individual case. The starting point is to express the collective spin in terms of bosonic modes: 
\begin{equation}
\hat{S}_+^{(j)}=\hat{b}_j^\dagger \sqrt{2S-\hat{b}_j^\dagger\hat{b}_j}, \quad  \hat{S}_-^{(j)}= \sqrt{2S-\hat{b}_j^\dagger\hat{b}_j}\,\,\hat{b}_j,
\end{equation}
with $j=1,2$. The bosonic modes are assumed in a large displaced state, around which we study the fluctuations:
\begin{equation}
\hat{b}_j\to\hat{b}_j+\sqrt{S}\beta_j,    
\end{equation}
where $\beta_j$ is a complex field to be determined. The  spin operators can be expanded in powers of the small parameter $\epsilon=1/\sqrt{S}$:
\begin{equation}
\hat{m}^{(j)}_\alpha=\frac{\hat{S}^{(j)}_\alpha}{S}=\sum_{l=0}^\infty \epsilon^l \hat{m}^{(j)}_{\alpha,l}.    
\end{equation}
Again, it is only relevant to write down the following explicit expressions:
\begin{equation}
{m}^{(j)}_{+,0}=\sqrt{k_j}\beta_j^*,\quad   \hat{m}^{(j)}_{+,1}=\frac{1}{2\sqrt{k_j}}\big[(4-3|\beta_j|^2)\hat{b}_j^\dagger-\beta_j^{*2}\hat{b}_j \big],  
\end{equation}
with $k_j=2-|\beta_j|^2$.

The expanded operators are to be used in the cascaded master equation with rescaled Rabi frequencies and time:
\begin{equation}
\partial_\tau\hat{\rho}=-iS\big[\tilde{\omega}\hat{m}^{(1)}_{x}+\tilde{\omega}_\mathrm{D}\hat{m}^{(2)}_{x},\hat{\rho}\big]
-\frac{\kappa}{2}S\big[\hat{m}^{(2)}_+\hat{m}^{(1)}_--\hat{m}^{(1)}_+\hat{m}^{(2)}_-,\hat{\rho}\big]
+\kappa S\mathcal{D}[\hat{m}^{(1)}_-+\hat{m}^{(2)}_-]\hat{\rho}.  
\end{equation}

The dominant term of order $\sqrt{S}$ yields self-consistent equations for $\beta_j$:
\begin{equation}
-i\tilde{\omega}\pm\kappa m_{\pm,0}^{(1)}=0,\quad
-i\frac{\tilde{\omega}_\mathrm{D}}{2}\pm\kappa m_{+,0}^{(1)}\pm\frac{\kappa}{2} m_{\pm,0}^{(2)}=0,
\end{equation}
which have as a solution
\begin{equation}
 m_{\pm,0}^{(1)}=\pm i\frac{\tilde{\omega}}{\kappa},\quad
    m_{\pm,0}^{(2)}=\mp i\frac{2\tilde{\omega}-\tilde{\omega}_\mathrm{D}}{\kappa},\quad
   \beta_1=-i\sqrt{1-\sqrt{1-\frac{\tilde{\omega}^2}{\kappa^2}}},\quad
   \beta_2=i\sqrt{1-\sqrt{1-\frac{(2\tilde{\omega}-\tilde{\omega}_\mathrm{D})^2}{\kappa^2}}},
\end{equation}
valid in the range: $\tilde{\omega}\leq \kappa$, and $|2\tilde{\omega}-\tilde{\omega}_\mathrm{D}|\leq \kappa$. This leads to the mean-field magnetizations in the stationary state with components (the x-components are zero):

\begin{equation}
\langle\hat{S}_\mathrm{y}^{(1)} \rangle_\mathrm{ss}=\frac{\omega}{\kappa}, \quad  \langle\hat{S}_\mathrm{z}^{(1)} \rangle_\mathrm{ss}=-\frac{N}{2}\sqrt{1-\frac{\omega^2}{\omega_c^2}},\quad 
\langle\hat{S}_\mathrm{y}^{(2)} \rangle_\mathrm{ss}=-\frac{2\omega-\omega_\mathrm{D}}{\kappa}, \quad  \langle\hat{S}_\mathrm{z}^{(2)} \rangle_\mathrm{ss}=-\frac{N}{2}\sqrt{1-\frac{(2\omega-\omega_\mathrm{D})^2}{\omega_c^2}}.
\end{equation}

The next leading term is a zero order term in $\epsilon$. This describes the leading dynamics of the fluctuations:
\begin{equation}
\partial_\tau\hat{\rho}_0=-\frac{\kappa}{2}\big[\hat{m}^{(2)}_{+,1}\hat{m}^{(1)}_{-,1}-\hat{m}^{(1)}_{+,1}\hat{m}^{(2)}_{-,1},\hat{\rho}_0\big]+\kappa \mathcal{D}[\hat{m}^{(1)}_{-,1}+\hat{m}^{(2)}_{-,1}]\hat{\rho}_0.  
\end{equation}
The stationary state for the fluctuations is the product state of vacuum states of each $\hat{m}^{(j)}_{-,1}$:
\begin{equation}
\hat{\rho}_{0,\mathrm{ss}}=\big(|E_0^{(1)}\rangle\otimes|E_0^{(2)}\rangle\big) \big(\langle E_0^{(1)}|\otimes\langle E_0^{(2)}|\big), \quad
\hat{m}^{(1)}_{-,1} |E^{(1)}_0\rangle=0\quad \hat{m}^{(2)}_{-,1} |E^{(2)}_0\rangle=0.
\end{equation}
The jump operators can be rewritten as:
\begin{equation}
\hat{m}^{(j)}_{-,1}=A_j\hat{b}_j+B_j\hat{b}_j^\dagger,    
\end{equation}
with $A_1=A$ and $B_1=B$ and:
\small
\begin{equation}
A_2=\frac{1+3\sqrt{1-(2\tilde{\omega}-\tilde{\omega}_\mathrm{D})^2/\kappa^2}}{2\sqrt{1+\sqrt{1-(2\tilde{\omega}-\tilde{\omega}_\mathrm{D})^2/\kappa^2}}}, \quad
B_2=\frac{1-\sqrt{1-(2\tilde{\omega}-\tilde{\omega}_\mathrm{D})^2/\kappa^2}}{2\sqrt{1+\sqrt{1-(2\tilde{\omega}-\tilde{\omega}_\mathrm{D})^2/\kappa^2}}}.
\end{equation}
\normalsize
These coefficients satisfy that $|B_j/A_j|\leq 1$ in the valid parameter regime. Then, the following state always exist in the stationary phase and makes up the vacuum state of the collective jump operator:
\begin{equation}
|E_0^{(j)}\rangle=\frac{1}{\sqrt{ \mathcal{N}_j}}|0\rangle_j+\frac{1}{\sqrt{ \mathcal{N}_j}}\sum_{n=1}^\infty(-1)^n\bigg(\frac{B_j}{A_j}\bigg)^n\sqrt{\frac{(2n-1)!!}{2n!!}}|2n\rangle_j,
\end{equation}
where $|n\rangle_j$ is a Fock state of boson $j$ and $\mathcal{N}_j$ is a normalization constant. This state has analogous properties as the one for the individual system.

\subsection{Calculation of the scaled cumulant generating function} 

The starting point is the tilted (rescaled) master equation for the cascaded system:
\begin{eqnarray}
\partial_\tau\hat{\rho}=-iS\big[\tilde{\omega}\hat{m}^{(1)}_{x}+\tilde{\omega}_\mathrm{D}\hat{m}^{(2)}_{x},\hat{\rho}\big]
-\frac{\kappa}{2}S\big[\hat{m}^{(2)}_+\hat{m}^{(1)}_--\hat{m}^{(1)}_+\hat{m}^{(2)}_-,\hat{\rho}\big]+\kappa S\mathcal{D}[\hat{m}^{(1)}_-+\hat{m}^{(2)}_-]\hat{\rho}\\ \nonumber
+\kappa S(e^{-s}-1)(\hat{m}_-^{(1)}+\hat{m}_-^{(2)})\hat{\rho}(\hat{m}_+^{(1)}+\hat{m}_+^{(2)})
\end{eqnarray}
We analyze this equation around the fixed point we have found previously and in the regime in which the fluctuation dynamics is well described by the flucutation operators of order $l=1$. Therefore, we plug in the fluctuation stationary state and the operators expanded up to this order in the tilted master equation. Integrating it for a small time step $\Delta\tau$ we obtain:
\begin{equation}
\hat{\rho}(\Delta \tau)\approx\hat{\rho}_{0,\mathrm{ss}}
+\Delta \tau \kappa S(e^{-s}-1)\big[{m}^{(1)}_{-,0}+{m}^{(2)}_{-,0}+\epsilon\hat{m}_{-,1}^{(1)}+\epsilon\hat{m}_{-,1}^{(2)}+\dots\big]\hat{\rho}_{0,\mathrm{ss}}[{m}^{(1)}_{+,0}+{m}^{(2)}_{+,0}+\epsilon\hat{m}_{+,1}^{(1)}+\epsilon\hat{m}_{+,1}^{(2)}\dots\big].
\end{equation}
Similarly to the individual system, the properties of $\hat{\rho}_{0,\mathrm{ss}}$ ensure that the leading contribution is of order $S$ and that there is no contribution of order $\sqrt{S}$ (as the stationary state is annihilated by the jump operator). Thus, up to order 1 the fluctuation stationary state is an eigenstate of the tilted master equation and we see that: 
\begin{equation}
\text{Tr}[\hat{\rho}(\tau)]\approx e^{\tilde{\theta}_c(s)\tau}, 
\end{equation}
where
\begin{equation}
\tilde{\theta}_c(s)=S(e^{-s}-1)\frac{(\tilde{\omega}-\tilde{\omega}_\mathrm{D})^2}{\kappa}  +\mathcal{O}[1],  
\end{equation}
is the (re-)scaled cumulant generating function. Taking into account that $\tau=St$, the scaled cumulant generating function for the bare time is related to the one for the rescaled time as:
\begin{equation}
\theta_c(s)=S \tilde{\theta}_c(s).    
\end{equation}
Thus, up to leading order we obtain:
\begin{equation}\label{SCGF_cascaded}
{\theta}_c(s)\approx(e^{-s}-1)\frac{(\omega-{\omega}_\mathrm{D})^2}{\kappa}.  
\end{equation}
Using the results of the Large deviations theory we obtain the estimation error presented in the main text:
\begin{equation}\label{error_analytical_cascaded}
\overline{\delta\omega}=   \frac{\sqrt{\theta_\mathrm{c}''(0)}}{|\partial_\omega \theta_\mathrm{c}' (0)|}\approx 0.5\sqrt{\kappa}.   
\end{equation}

\begin{figure}[t!]
 \centering
 \includegraphics[width=0.6\columnwidth]{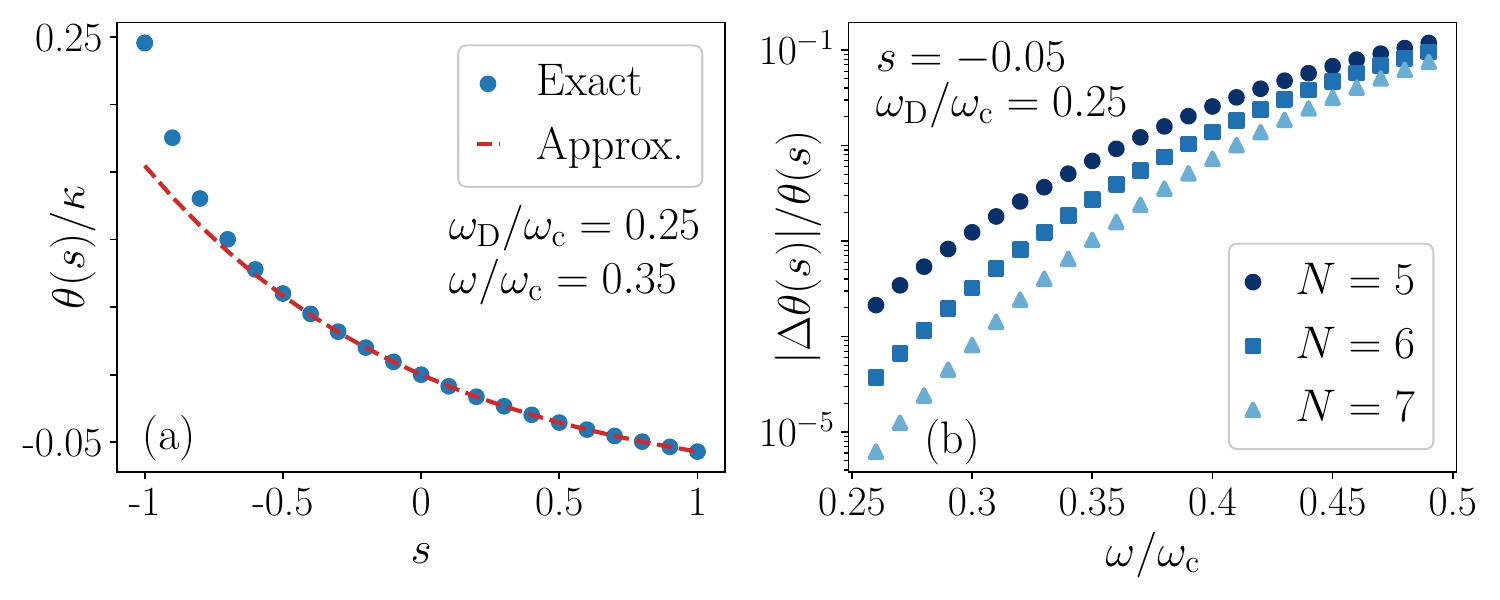}
 \caption{{\bf Benchmark  for the scaled cumulant generating function in the cascaded system.} (a) Scaled cumulant generating function [Eq. (\ref{SCGF})] varying $s$ for $\omega/\omega_\mathrm{c}=0.35$, $\omega_\mathrm{D}/\omega_\mathrm{c}=0.25$ and $N=6$. The blue points are obtained diagonalizing the tilted master equation. The red dashed line corresponds to the analytical result of Eq. (\ref{SCGF_cascaded}). (b) Relative error on $\theta(s)$ as a function of $\omega/\omega_\mathrm{c}$, for different system sizes, $\omega_\mathrm{D}/\omega_\mathrm{c}=0.25$ and $s=-0.05$. This error is defined as the difference between the exact result and the analytical result of  Eq. (\ref{SCGF_individual}), both divided by the exact result.}
 \label{fig_SCGF_cascaded}
\end{figure}

{\it Numerical benchmark. --} In Fig. \ref{fig_SCGF_cascaded}, we compare the analytic approximation [Eq. (\ref{SCGF_cascaded})] with the result obtained numerically diagonalizing tilted master equation. We find good agreement, though errors are larger than in the individual case. We observe that the relative error [Fig. \ref{fig_SCGF_cascaded} (b)] increases near the transition, however it is still generally smaller than $1/N$.

\section{Separation of timescales between collective and local processes}

Here we discuss the separation of timescales between the collective time-crystal oscillations and possible local decay channels. For large atom numbers and the system in the time-crystal phase, the oscillations display a dominant frequency close to the mean-field one: $\Omega\approx \frac{N\kappa}{2}\sqrt{(\omega/\omega_\mathrm{c})^2-1}$ \cite{Cabot2022,Carmichael1980}. We now assume  that besides collective dissipation there are also moderate local dissipation and dephasing processes occurring with a rate $\gamma_\mathrm{loc}$. The characteristic timescale associated to these processes is $T_\mathrm{loc}=\gamma^{-1}_\mathrm{loc}$. On the other hand, the characteristic timescale associated to the collective oscillations is their period: $T_\mathrm{osc}=2\pi/\Omega\sim N^{-1}$. For large enough atom numbers we  have $T_\mathrm{osc}\ll T_\mathrm{loc}$, i.e., a large separation of timescales. This means that, when initially preparing  the ensemble in a fully symmetric state  (e.g. all atoms excited), many time-crystal oscillations will occur before the effects of local dissipation are significant. These local incoherent terms break the conservation of total angular momentum and we expect the system to reach a different stationary state \cite{Piccitto2021}. However, when  $T_\mathrm{osc}\ll T_\mathrm{loc}$, we expect the time-crystal oscillations to manifest as a metastable phenomenon \cite{Macieszczak2016b,Zhu2019,Cabot2022b} in which the characteristic separation of timescales increases with atom number.

\section{Error bars of the Montecarlo results for the cascaded system}\label{app_error}

In the study of the sensitivity for the cascaded system, we have used quantum trajectories to evaluate the estimation error for large system sizes. In this regard, we distinguish between three subsets of data:
\begin{itemize}
    \item For $N\leq 11$, both the prefactor of the standard deviation of the intensity, $\overline{\sigma_{I_\mathrm{T}}}=\sqrt{(\mathbb{E}[I^2_T]-\mathbb{E}[I_T]^2)T}$, and the derivative of the intensity, $\partial_\omega I_\mathrm{T}$, have been obtained from eigenvalue methods. These consist on finding the largest eigenvalue of the tilted master equation or finding the stationary state of the master equation and performing a numerical derivative, respectively.
    \item For $11<N\leq 22$, $\overline{\sigma_{I_\mathrm{T}}}$ has been obtained from Montecarlo sampling of quantum trajectories, while $\partial_\omega I_\mathrm{T}$ from eigenvalue methods.
    \item For $N>22$, only Monte Carlo sampling of quantum trajectories is available due to its more favourable scaling in computational cost. In this case, both quantities have been obtained from quantum trajectories.
\end{itemize}

In Fig. 4 (d) of the main text, we have introduced error bars  in order to account for errors associated to the Montecarlo methods. We account for different source of errors:  (i) Montecarlo error due to finite sampling; (ii) systematic errors due to finite measurement time runs and effect of initial conditions.  For system sizes in the range $11<N\leq 22$, we have used samples of $10^5$ trajectories per point. For system sizes in the range $N>22$, we have used larger samples of $5\cdot10^5$ trajectories. This is because in order to perform  the numerical derivative of $\partial_\omega I_\mathrm{T}$ the quantity $I_\mathrm{T}$ needs to be determined with high precision (a step $\kappa d\omega=0.00025$ is used to ensure enough precision in the numerical derivative). In both cases, quantum trajectories are computed for a total measurement time $\kappa T=440$. Simulations are performed such that data is started to be compiled after $\kappa t=40$, in order to minimize the effects of initial conditions. Then, we write the standard deviation on the estimation error as (error propagation formula):
\begin{equation}
\text{Err}\big[\overline{\delta\omega}\big]\approx\frac{\text{Err}\big[(\mathbb{E}[I^2_T]-\mathbb{E}[I_T]^2)T\big]}{2\overline{\sigma_{I_T}}|\partial_{\omega}I_T|}+\frac{\overline{\sigma_{I_T}}}{|\partial_{\omega}I_T|^2}\text{Err}\big[|\partial_{\omega}I_T|\big].
\end{equation}
Comparing Montecarlo results with the ones obtained from eigenvalue methods (for sizes in which it is possible to do so), we observe errors of the order of $1-2\%$ in $(\mathbb{E}[I^2_T]-\mathbb{E}[I_T]^2)T$ and errors of the order of $5\%$ in $\partial_\omega I_\mathrm{T}$. We thus make the estimation that Err$[(\mathbb{E}[I^2_T]-\mathbb{E}[I_T]^2)T]\sim 0.02(\mathbb{E}[I^2_T]-\mathbb{E}[I_T]^2)T$ and  Err$[|\partial_{\omega}I_T|]\sim0.05|\partial_{\omega}I_T|$.

\end{document}